\documentclass[a4paper,11pt]{article}

\usepackage{physics}
\usepackage[usenames,dvipsnames]{xcolor}
\usepackage{caption}
\usepackage{subcaption}
\usepackage{siunitx}
\usepackage{booktabs}
\usepackage{pdfcolfoot}

\pdfoutput=1 % if your are submitting a pdflatex (i.e. if you have
\usepackage{jheppub} % for details on the use of the package, please
\hypersetup{%
  colorlinks=true, linktocpage=true, pdfstartpage=1, pdfstartview=FitV,
  breaklinks=true, pageanchor=true,
  pdfpagemode=UseNone,
  plainpages=false, bookmarksnumbered, bookmarksopen=true, bookmarksopenlevel=1,
  hypertexnames=true, pdfhighlight=/O,
  urlcolor=RoyalBlue, linkcolor=ForestGreen, citecolor=RoyalBlue,
  pdftitle={\boldmath $B_s \to \mu^+ \mu^-$ in a Two-Higgs-Doublet Model with flavour-changing up-type Yukawa couplings},
  pdfauthor={Martin S.\ Lang, Ulrich Nierste},
  pdfsubject={},
  pdfkeywords={},
  pdfcreator={pdfLaTeX},
  pdfproducer={LaTeX with hyperref}
}

\newcommand{\hc}{h.c.}
\newcommand{\no}{\nonumber}
\newcommand{\nn}{\nonumber \\}
\newcommand{\lt}{\left}
\newcommand{\rt}{\right}
\newcommand{\gev}{\,\mbox{GeV}}

\newcommand{\fig}[1]{Fig.~\ref{#1}}
\newcommand{\tab}[1]{Tab.~\ref{#1}}
\newcommand{\eq}[1]{Eq.~(\ref{#1})}
\newcommand{\eqsand}[2]{Eqs.~(\ref{#1}) and (\ref{#2})}
\newcommand{\eqsto}[2]{Eqs.~(\ref{#1}--\ref{#2})}
\newcommand{\ov}{\overline}
\newcommand{\mx}[1]{\begin{pmatrix} #1 \end{pmatrix}}
\newcommand{\ds}{\displaystyle}
\newcommand{\lag}{\ensuremath{\mathcal{L}}}

\author{Martin S.\ Lang,}
\author{Ulrich Nierste}

\affiliation{Institute for Theoretical Particle Physics, Karlsruhe Institute of Technology (KIT),\\
  Wolfgang-Gaede-Str. 1, D-76131 Karlsruhe, Germany}

\emailAdd{m.lang@kit.edu}
\emailAdd{ulrich.nierste@kit.edu}

\abstract{We present a Two-Higgs-Doublet Model in which the structure of
  the quark Yukawa {matrices} is governed by three spurions breaking
  the flavour symmetries {of the quark Yukawa sector}. The model
  naturally suppresses {flavour-changing neutral current (FCNC)}
  amplitudes in the down-type sector, but permits sizable FCNC couplings
  in the up sector. We calculate the branching ratio of
  $B_s \to \mu^+ \mu^-$ to leading and next-to-leading order of QCD for
  the case with FCNC couplings of the heavy neutral Higgs bosons to up-type 
  quarks and verify that all counterterms follow the pattern
  dictated by the spurion expansion of the Yukawa matrices.  We find
  correlations between $B_s \to \mu^+ \mu^-$, $b\to s\gamma$, and the
  Higgs masses. The $B_s - \bar B_s$ mixing amplitude is naturally
  suppressed in the model but can probe a portion of the parameter space
  with very heavy Higgs bosons.}

\title{\boldmath $B_s \to \mu^+ \mu^-$ in a Two-Higgs-Doublet Model with flavour-changing up-type Yukawa couplings}

\begin{document}

\maketitle
\flushbottom

%- {{{ Introduction:

\section{Introduction}\label{sec:intro}

Two-Higgs-Doublet models (2HDMs)~\cite{Branco:2011iw,Gunion:1989we} are
popular extensions of the Standard Model (SM) due to their relative
simplicity, involving no additional fields apart from a second
Higgs doublet.  Moreover, a strong motivation to study 2HDMs also comes
from theories in which a second Higgs doublet is required due to
symmetry arguments, e.g.\ axion models in the context of the strong CP
puzzle~\cite{Kim:1986ax,Peccei:1977hh,Peccei:1977ur} or minimal
supersymmetry~\cite{Haber:1984rc}. 2HDMs differ in the structure of
Higgs-fermion Yukawa couplings. The historically most favoured variants
{are the so-called type-I and type-II} 2HDM in which {both} up-type and
down-type quarks {only} couple to {one of the two} Higgs doublets.  In
{these types of} 2HDMs, flavour-changing neutral current processes
{(FCNC)} such as the decay $B_s \to \mu^+ \mu^-$ are loop-suppressed and
therefore small masses of the additional Higgs bosons are in principle
possible, an appealing feature {in the time of} early LHC searches.
The generic 2HDM (also dubbed ``type-III''), with most general Yukawa
  matrices, exhibits, however, a much richer {phenomenology}
\cite{Crivellin:2012ye,Crivellin:2013wna,Crivellin:2019dun} in which
flavour-changing neutral Higgs-boson couplings are possible, in case
up-type or down-type quarks couple to more than one Higgs
doublet.  The generic 2HDM suffers from a low level of predictivity
  caused by its large number of parameters and moreover requires that
  FCNC couplings in the down-type sector are tuned to small values in an
  ad-hoc way to comply with the many experimental constraints from
  $s\to d$, $b\to s$, and $b\to d$ FCNC processes. Yet the economic 2HDM
  of type I and II cannot address current anomalies in rare bottom quark
  decays, for instance data favour an excess of 
  $\mathcal{B} (B\to D \tau \nu)/\mathcal{B} (B\to D \mu \nu)$ over its
  SM prediction \cite{HFLAV:2022pwe,BaBar:2012obs,BaBar:2013mob,Belle:2015qfa,LHCb:2015gmp,Belle:2016dyj,Belle:2017ilt,LHCb:2017smo,LHCb:2017rln,Belle:2019rba,LHCb:2023zxo,LHCb:2023uiv} and this
  cannot be explained in
  these models. Also to address the observed deficit in
  $b\to s\mu^+\mu^-$ for low values of the lepton-pair invariant mass
  one resorts to the generic 2HDM \cite{Crivellin:2019dun}.

In this paper, we propose a three-spurion 2HDM, a 2HDM which is more general than the
  popular type-I and type-II models (which it contains as limiting
  cases) but more constrained than the generic 2HDM.
  A process on which
  our model has characteristic imprints is $B_s \to \mu^+ \mu^-$, which
  we study in detail in this paper, including the calculation of
  two-loop QCD corrections.  The model under consideration will feature
flavour-changing up-type Yukawa couplings of the additional {heavy}
neutral Higgs bosons, most notably a charm-top transition which may
eliminate the Cabibbo-Kobayashi-Maskawa (CKM) suppression {in $b\to s$
transitions} that is present in both the SM and the 2HDM type-II models.

The outline of the paper is as follows: The Yukawa and Higgs sectors of the three-spurion 2HDM
are presented in \autoref{sec:yukawasector}, with particular emphasis on some peculiarities of this specific model.
In \autoref{sec:bsmumu}, we will
introduce the effective operators contributing to the low-energy weak
$B_s \to \mu^+ \mu^-$ decay, while \autoref{sec:setup}
is dedicated to a short description of the computational setup used for
the evaluation of Feynman diagrams. The limiting case of the type-II 2HDM
is discussed in \autoref{sec:decay2hdm}, with the additional
contributions from flavour-changing neutral-Higgs Yukawa couplings being
presented in \autoref{sec:newcontributions}.
Finally, we will discuss
the phenomenology of such models in \autoref{sec:phenopart} and summarize in \autoref{sec:summary}.
%- }}}
%- {{{ Yukawa sector of the 2HDM:

\section{2HDM with suppressed down-type FCNC
  couplings}\label{sec:yukawasector}
In this section we introduce the so-called \emph{three-spurion 2HDM}, which allows for significant flavour-changing Yukawa couplings in the up-type quark sector, while the ones in the down-type quark sector are naturally suppressed.

\subsection{General Yukawa sector}
Our starting point, the quark Yukawa Lagrangian of the general (``type III'') 2HDM, reads
\begin{align}
L_Y &=  -\ov{Q}_f^\prime \Big[ \ov Y_{fi}^d H_d + \ov\epsilon_{fi}^d H_u 
            \Big]d_{iR}^\prime 
        \;-\; \ov{Q}_f^\prime  \Big[ \ov Y_{fi}^u \epsilon H_u^* 
           + \ov\epsilon_{fi}^u \epsilon H_d^*  \Big]u_{iR}^\prime  +\hc \nn
&\equiv -\ov{Q}^\prime \Big[ \ov Y^d H_d + \ov\epsilon^d H_u \Big]d_{R}^\prime  
        \;-\;\ov{Q}^\prime \Big[ \ov Y^u \epsilon H_u^* 
           + \ov\epsilon^u \epsilon H_d^*  \Big]u_{R}^\prime  +\hc
\label{ly}
\end{align}
with four general complex $3\times 3$ matrices $\ov Y^{u,d}$
  and $\ov\epsilon^{u,d}$ as well as
\begin{align}
  H_{u,d} &= \begin{pmatrix} H_{u,d}^+ \\ H_{u,d}^0\end{pmatrix}, 
   & 
 \epsilon H_{u,d}^* &= 
  \begin{pmatrix} H_{u,d}^{0*} \\ -H_{u,d}^- \end{pmatrix}, \\
 \mbox{and~~~}Q_f^\prime &= 
  \begin{pmatrix} u_{fL}^\prime \\ d_{fL}^\prime \end{pmatrix} .
\end{align}
The subscripts $f,i=1,2,3$ label the generations, e.g.\ $u_{3L}^\prime=t_{L}^\prime$.
The notation of \eq{ly} follows Ref.~\cite{Crivellin:2013wna},
except that our $H_d$ corresponds to $-\epsilon H_d^*$ of that paper.
The vacuum expectation values (vevs) and related quantities are
\begin{align}\label{vevs}
\langle H_u^0\rangle & = v_u = v\sin \beta, &  
\langle H_d^0\rangle & = v_d = v\cos \beta, \no\\[2mm]
 \tan\beta & := \frac{v_u}{v_d}, & 
 v&:= \sqrt{v_u^2+v_d^2} = 174\,\mbox{GeV}  .
\end{align}
The quark mass matrices are 
\begin{align}\label{mdu}
M^d &= \ov Y^d v \cos \beta + \ov\epsilon^d v \sin\beta,
 & 
M^u &= \ov Y^u v \sin \beta + \ov\epsilon^u v \cos\beta,
\end{align}
which we 
diagonalise in the usual
way with the help of unitary matrices $S_{L,R}^{u,d}$,
\begin{align}
u_{L,R}^\prime = S_{L,R}^u u_{L,R},&   &   
d_{L,R}^\prime = S_{L,R}^d d_{L,R}, & \label{uup}\\
S_L^{d\dagger} M^d S_R^d = \hat M^d =& 
\mx{m_d & 0 & 0\\ 0 & m_s & 0 \\ 0 & 0 & m_b}, & 
S_L^{u\dagger} M^u S_R^u = \hat M^u =& 
\mx{m_u & 0 & 0\\ 0 & m_c & 0 \\ 0 & 0 & m_t}, \label{wm}
\end{align}
with the unprimed fields corresponding to quark mass eigenstates.
The gauge sector of the 2HDM is invariant under independent unitary rotations of the
fields $Q^\prime$, $d_{R}^\prime$, and $u_{R}^\prime$ in flavour
space. We use \eq{uup} and choose
\begin{align}
  Q^\prime & \equiv   S_L^d Q \label{qqprime}  
\end{align}
and find $L_Y$ in the so-called \emph{down basis}:
\begin{align}
L_Y &\equiv -\ov{Q} \Big[  Y^d H_d + \epsilon^d H_u \Big]d_{R} 
        \;-\;\ov{Q}  \, \Big[  Y^u \epsilon H_u^* 
           + \epsilon^u \epsilon H_d^*  \Big]u_{R}  +\hc
\label{lyd}
\end{align}
with the appropriately transformed Yukawa matrices
\begin{align}
Y^{u,d} &= S_L^{d\dagger} \ov Y^{u,d} S_R^{u,d}, & 
\epsilon^{u,d} &= S_L^{d\dagger} \ov \epsilon^{u,d} S_R^{u,d}  
\end{align}
and the CKM matrix
\begin{align}
V &= S_L^{u\dagger} S_L^d. \label{ckmm}
\end{align}
The Yukawa Lagrangian $L_Y$ in \eq{lyd} is manifestly SU(2) invariant, with the SU(2) doublet
\begin{align}
 Q &=  S_L^{d\dagger} Q^\prime = \mx{V^\dagger \,u_L \\ d_L} .\no
\end{align}
\eq{lyd} is our starting point; the Yukawa matrices are related
to the diagonal mass matrices as
\begin{align}\label{mdum}
\frac{\hat M_d}{v} &=\, Y^{d} \cos \beta + \epsilon^d \sin\beta, & 
\frac{\hat M_u}{v} &=\, V\lt( Y^{u} \sin \beta + \epsilon^u \cos\beta \rt).
\end{align}
Non-vanishing  off-diagonal entries of $Y^{d,u}$, $\epsilon^{d,u}$ give
rise to FCNC couplings of the neutral components of the Higgs doublets. 

In a general 2HDM the quantity $\tan \beta$ has no physical meaning: One
can arbitrarily rotate $\mx{H_u,H_d}$ in \eq{ly} leading to a Lagrangian
of the same form (yet with different Yukawa matrices) and the rotation
angle will add to $\beta$. The situation is different in variants of the
2HDM in which $H^u$ and $H^d$ are distinguished by quantum numbers which
forbid such rotations of $\mx{H_u,H_d}$. 
Prominent examples are the 2HDM of type I and II, in which two out
of the four Yukawa matrices in \eqsand{ly}{lyd} are absent. The type-I
model corresponds to $\epsilon^d=Y^u =0$. The type-II model is instead
found for $\epsilon^d=\epsilon^u =0$.

The doublets $\phi_{\rm SM}$, $\phi_{\rm new}$ of the \emph{Higgs basis}
\cite{Haber:2006ue,Eriksson:2009ws,Branco:1999fs}
are defined by a rotation of the two doublets $H_u$ and $H_d$ 
by the angle $\beta$ 
such that $\phi_{\rm new}$ has no vev:
\begin{align}\label{hb}
\mx{\phi_{\rm new}\\\phi_{\rm SM} } &=
\mx{\cos\beta & -\sin\beta\\ \sin\beta & \phantom{-}\cos\beta} \, 
\mx{H_u \\ H_d}
\end{align}
One has
\begin{align}\label{smnew}
\phi_{\rm SM}&= \mx{\ds G^+\\[1mm] \ds v+\frac{\phi^0+i G^0}{\sqrt{2}} }, &
\phi_{\rm new}&= \mx{\ds H^+\\[1mm] \ds \frac{\phi^{0\prime}+i
  A^0}{\sqrt{2}} } \,.%
\end{align}
Next we use \eq{hb} to express $L_Y $ in terms of $\phi_{\rm SM}$ and $ \phi_{\rm new}$:
\begin{align}\label{lysn}
L_Y =& \; \quad \ov{Q} \Big[ - Y^d \cos \beta - 
                        \epsilon^d \sin\beta 
                \Big] \phi_{\rm SM} d_{R}
         \quad\,+\,\ov{Q}  \Big[   Y^d \sin \beta - 
                        \epsilon^d \cos\beta 
                \Big] \phi_{\rm new} d_{R}
\nn
& + \ov{Q} V^\dagger \Big[ - Y^u \sin \beta -
                        \epsilon^u \cos\beta
                \Big] \epsilon \phi_{\rm SM}^* u_{R} \nn
&  +\,    \ov{Q} V^\dagger \Big[  - Y^u \cos \beta +
                        \epsilon^u \sin\beta
                        \Big] \epsilon \phi_{\rm new}^* u_{R}
 +\hc
\end{align}

By using \eq{mdum} to eliminate $Y^u$ and $Y^d$ one can write the
couplings to the physical charged and neutral Higgs bosons in \eq{lysn} as
\begin{align}\label{eq:yuklag}
 \lag_{Y}^{\rm phys} = \, & -\bar{u}_{L} \left[\frac{ \hat{M}^u}{v}
                    \cot\beta +
                    g^u \right] u_{R} \frac{\phi^{0\prime} - \mathrm{i} A^0}{\sqrt{2}} 
                    + \bar{d}_L \left[ \frac{\hat{M}^d}{v} \tan
                    \beta
                    + g^d \right] d_{R} \frac{\phi^{0\prime} + \mathrm{i} A^0}{\sqrt{2}} \notag \\
                  & + \bar{u}_{L} V \left[ \frac{\hat{M}^d}{v}
                    \tan\beta +
                    g^d \right] d_{R} H^+ + \bar{d}_L V^\dagger
                    \left[ \frac{ \hat{M}^u}{v} \cot\beta + g^u \right]
                    u_R H^- \nn
                  & -\ov{d}_L \frac{\hat M^d}{v} d_{R} \,  
   \lt( v + \frac{\phi^0}{\sqrt{2}} \rt)
       \; - \;
             \ov{u}_L \frac{\hat M^u}{v} u_{R} \,
   \lt( v + \frac{\phi^0}{\sqrt{2}} \rt) \;+\; \hc 
\end{align}
with the matrices \cite{Crivellin:2013wna}
\begin{align}\label{adu}
g^d &= -\epsilon^d \sin\beta \lt( \tan\beta +\cot \beta \rt), & 
g^u &= -\epsilon^u \cos\beta \lt( \tan\beta +\cot \beta \rt).  
\end{align} 
The non-diagonal matrices $g^d$ and $g^u$ characterise the deviations
from the popular type-II 2HDM (for which $g^d=g^u=0$) and can induce
flavour-changing couplings of neutral Higgs bosons.  Note that the
type-I model is also included in the formalism and recovered by using
\eq{mdum} with $Y^u=0$ in the expression for $g^d$. For our loop
calculation and the phenomenological analysis it is advantageous to work
with $g^{d,u}$ rather than $\epsilon^{d,u}$, especially for the
definition of the renormalisation prescriptions.  We write
$g_{d_i d_j} \equiv g^d_{ij}$, where $d_i$ is the $i$-th down-type quark
flavor, $i=1,2,3$, and similarly for $g^u$.

We restrict 
ourselves to the CP-conserving Higgs potential, such that
$A^0$ is a pseudoscalar boson, while $\phi^0$ and $\phi^{0\prime}$
are scalar particles.
The two Higgs bosons $\phi^0$ and $\phi^{0\prime}$ are in
general not mass eigenstates. The latter are given by
$h^0$ and $H^0$, with
\begin{equation}
  \begin{pmatrix} h^0 \\ H^0 \\ A^0 \end{pmatrix} =
  \begin{pmatrix} \sin\left(\beta - \alpha \right) & \cos\left(\beta -
      \alpha \right) & 0 \\
    \cos\left(\beta - \alpha \right) & -\sin\left(\beta - \alpha \right)
    & 0 \\ 0 & 0 & 1
  \end{pmatrix} \,
  \begin{pmatrix} \phi^0 \\ \phi^{0\prime} \\ A^0 \end{pmatrix} \,.
  \label{hmix}
\end{equation}
The angle $\alpha$ is a priori arbitrary, but data on decays of the
  125\gev\ Higgs boson constrain $\cos(\beta -
      \alpha )$ to be close to zero.

\subsection{Spurion expansion}
The 2HDM of type I and II (and their variants with
modified lepton couplings) invoke (softly broken) $\mathbb{Z}_2$
symmetries to forbid FCNC couplings of the neutral Higgs bosons. This
was motivated by the wish to find non-standard Higgs bosons at modern
colliders, because for generic values of the Yukawa matrices in \eq{lyd}
constraints from FCNC processes like meson-antimeson mixing push these
masses to values outside the reach of LEP, Tevatron, and LHC. With
the absence of discoveries of non-standard Higgs bosons this line of
arguments loses its appeal and the consideration of more general Yukawa  
sectors is in order.

The type-II model is the most studied variant of the 2HDM for two
reasons: Firstly, it constitutes the tree-level Higgs sector of the
Minimal Supersymmetric Standard Model (MSSM), in which the holomorphy of
the superpotential enforces $\epsilon^{u,d}=0$. Secondly, the type-II
model is phenomenologically especially interesting, because in this
model FCNC processes are sensitive to loop effects of the charged Higgs
boson.  A prominent example of the latter feature is the branching ratio $\mathcal{B}(B\to s\gamma)$,
which sets a stringent bound on the charged-Higgs mass
\cite{Misiak:2017bgg}. The type-II 2HDM further permits the possibility
of large down-type Yukawa couplings, a scenario motivated by the
possibility of bottom-top Yukawa unification. Such large-$\tan\beta$
scenarios are efficiently constrained by $\mathcal{B}(B_s\to \mu^+\mu^-)$
\cite{Logan:2000iv,Bobeth:2013uxa,Hermann:2013kca} and we will come back
to this topic in \autoref{sec:decay2hdm}. Concerning the first
above-mentioned motivation, the Higgs sector of the MSSM is really
richer than that of the type-II 2HDM: In the limit of infinitely heavy
superpartners one encounters non-decoupling loop-induced Yukawa matrices
$\epsilon^{u,d}$, an effect caused by the supersymmetry-breaking terms
\cite{Banks:1987iu,Hall:1993gn,Carena:1994bv}.  Despite the loop
suppression large effects are possible in FCNC processes with down-type
quarks which involve the product $\epsilon^d \tan\beta$
\cite{Babu:1999hn,Isidori:2001fv,Buras:2002vd,
  Buras:2002wq,Hofer:2009xb,Gorbahn:2009pp} with huge impact on
$\mathcal{B}(B_s\to \mu^+\mu^-)$ \cite{Babu:1999hn,Dedes:2001fv,Isidori:2001fv,Buras:2002wq,Buras:2002vd}.

The phenomenological constraints from meson-antimeson mixing and rare
(semi-)\-lep\-tonic decays place severe bounds on the off-diagonal
elements of $\epsilon^d $, while those of $\epsilon^u$ are essentially unconstrained
except for $\epsilon^u_{12}$ and $\epsilon^u_{21}$.  This situation
calls for a variant of the general 2HDM in which $Y^{u,d}$ and $\epsilon^u$ are
arbitrary, while $\epsilon^d$ is suppressed.  A strong motivation for
such a model is the possibility of spontaneous CP violation, implemented
through a Higgs potential developing complex vevs and real Yukawa
matrices. Spontaneous CP violation is not possible with a 2HDM of type I
or II, but requires at least three out of the four matrices in \eq{lyd}
to be non-zero. Avoiding fine-tuning implies that the dominant piece  of the
needed effect stems from $\epsilon^u$, while $\epsilon^d$ can be
neglected \cite{Nierste:2019fbx}.  However, in such a model, the mixing
of the neutral Higgs fields is different from \eq{hmix} and instead
involves all three fields. Yet for the CP-conserving observables
considered in this paper this feature is of minor relevance. The 2HDM
scenario with sizable $Y^{u,d}$ and $\epsilon^{u}$ has the appealing
feature that it simultaneously permits both measurable effects in FCNC
processes \emph{and}\ sufficiently light masses of the
non-standard Higgs particles enabling their discovery at the LHC.

Setting $\epsilon^d$ naively to zero leads to a non-renormalisable
model, because there are UV-divergent loops involving up-type quarks
with $Y^{u,d}$ and $\epsilon^{u}$ couplings, which require counterterms
proportional to elements of $\epsilon^d$. Whenever one seeks to
constrain the elements $g^u_{jk}$ of \eq{eq:yuklag} from FCNC
transitions of down-type quarks, one must foresee such a counterterm to
find a meaningful prediction. For example, $B_s\to \mu^+\mu^-$ is a
$b\to s $ transition constraining $g_{ct}$ and the corresponding loop
contribution requires a counterterm for $g_{sb}$. The minimal
renormalisable theory is found by invoking flavour symmetries and
systematically expanding in terms of the spurions breaking these
symmetries \cite{Chivukula:1987py,DAmbrosio:2002vsn}. The 2HDM gauge
sector is invariant under unitary rotations $Q\to U_QQ $,
$d_R\to U_dd_R$, $u_R\to U_uu_R$ in quark flavour space with
$(U_Q,U_d,U_u) \in SU(3)^3$ and the Yukawa sector is formally invariant
under this flavour symmetry if one transforms the matrices in \eq{lyd}
as
\begin{align}
  Y^{u,d} \to   U_Q Y^{u,d} U_{u,d}^\dagger , \qquad \qquad 
  \epsilon^{u,d} \to   U_Q \epsilon^{u,d}  U_{u,d}^\dagger. \label{flsym}
\end{align}

We propose to categorise the classes of renormalisable 2HDM in terms of
the spurions breaking the $ SU(3)^3$ flavour symmetry of the quark
sector.\footnote{The generalisation to the lepton sector is
  straightforward, but not relevant for the calculations in this paper.}
The minimal choice are two spurions, with just two physically distinct
possibilities. Both comply with the definition of minimal flavour
violation (MFV) as defined in Ref.~\cite{DAmbrosio:2002vsn}. The first
possibility is to take $Y^{u,d}$ as spurions and express the other two Yukawa
matrices as
$\epsilon^{u,d} = P_{u,d} (Y^{u}Y^{u\dagger}, Y^{d}Y^{d\dagger})\,
Y^{u,d}$, where $P_{u}$ and $P_{d}$ are polynomials. This variant is
discussed in Ref.~\cite{DAmbrosio:2002vsn} and amounts to a
generalisation of the 2HDM of type II. It also constitutes a
renormalisable extension of the aligned 2HDM of
Ref.~\cite{Tuzon:2010vt,Celis:2012dk,Eberhardt:2020dat}, in
which $P_{u,d}$ are constants.
If the 2HDM is the low-energy limit of a more fundamental
theory obeying the considered two-spurion symmetry-breaking pattern, the
latter will naturally induce $\epsilon^{u,d}$ in the described way as
well. An example for such a UV theory is the MSSM with a flavour-blind
supersymmetry breaking mechanism (such as gauge mediation). The second
possibility is to choose $Y^d$, $\epsilon^u$ as spurions, leading to a
generalisation of the type-I 2HDM.

There are two possibilities for a 2HDM with three spurions,
which can be taken as $Y^{u,d}$ plus either $\epsilon^u$ or $\epsilon^d$.
The first possibility is the phenomenologically interesting one and
studied in this paper. The expansion up to third order reads
\begin{align}
  \epsilon^d =&\; c Y^d \,+\, c_{11}   Y^d Y^{d\dagger} Y^d \nn
               & \phantom{\; c Y^d } \,+\, 
              b_{11}   \epsilon^u \epsilon^{u\dagger} Y^d 
              \,+\,  b_{12}   \epsilon^u Y^{u\dagger} Y^d \,+\,  
                           b_{21}   Y^u \epsilon^{u\dagger} Y^d
             \,+\,  b_{22}   Y^u Y^{u\dagger} Y^d  \label{spu}
\end{align}
with complex coefficients $c, \ldots, b_{22}$. 

Concerning \eq{spu} several comments are in order:
\begin{itemize}
\item The spurion expansion is only meaningful, if the contributions to
  the off-diagonal elements of $\epsilon^d$ from higher electroweak
  orders (with five or more Yukawa matrices) are small, so that they can
  be neglected. A  sufficient condition for this is realised in
  scenarios in which $c_{11},\ldots, b_{22}$  are induced by one-loop
  contributions in either the UV completion  or the 2HDM, while
  terms with $(2n+1)$ spurions are only generated at $n$-loop order and beyond.
  We consider this scenario throughout this paper.\footnote{A different
    application of \eq{spu}, in which $c_{11},\ldots, b_{22}={\cal
      O}(1)$ is allowed, is the case that $Y^u$ and $\epsilon^u$ are
    almost aligned, so that \eq{mdum} means small
    off-diagonal matrix elements  of these matrices in the chosen basis. Since $Y^u_{33},
    \epsilon^u_{33}={\cal O}(1)$ and  further $Y^d_{33}={\cal O}(1)$ is
    possible, some terms with five spurions must be added to \eq{spu},
    just as in the MFV  case of Ref.~\cite{DAmbrosio:2002vsn}.}
  Additional QCD corrections (e.g.\ an extra loop with a gluon) comply
  with the pattern in \eq{spu}, i.e.\ QCD renormalises the coefficients,
  but does not induce new ones. 
\item By rotating $(H_u,H_d)$ in \eq{lyd} one can eliminate $c$ in
  \eq{spu}. But in general radiative corrections bring this term back and
  a counterterm to $c$ is needed, unless one corrects the rotation in
  each order of perturbation theory. It is therefore advisable to keep
  $c$ in \eq{spu}; we treat it as a perturbative quantity with $c=0$ at tree level.
\item
  The decay $B_s\to \mu^+\mu^-$, which is the focus of the
    phenomenological analysis in this paper, involves 
  the FCNC vertex $\ov{Q}_j$-$d_{Rk}$-$H_d$ with $(j,k)=(2,3)$. The
  dominant one-loop  vertex diagram involves an internal $H_d$ line 
  and the product $Y^dY^{d\dagger} Y^d $ or
  $\epsilon^u\epsilon^{u\dagger} Y^d $ stemming from the three $H_d$ Yukawa
  couplings.
  The UV divergences
  can be cancelled by counterterms to $c_{11}$ and $b_{11}$.
 \item With \eq{mdum} we can trade $Y^{u,d}$ in \eq{spu} for the
  quark masses and CKM elements. Compared to the SM we find 14
  additional complex parameters, the 9 entries of $\epsilon^u$ and
  $c_{11},\ldots b_{22}$. Yet it is much more convenient to
  express observables in terms of  $g_{jk}^u$ of \eq{eq:yuklag} instead
  of $\epsilon^u_{jk}$ 
  and then use \eqsand{adu}{spu} (with \eq{mdum}) to calculate the
  $g_{jk}^d$ in terms of the coefficients of the spurion expansion.
  While this procedure is needed in 
  a global analysis of all available data ---which is beyond
  the scope of this paper---, the study of $b\to s$ transitions alone
  will simply involve $g_{ct}$ and, with CKM suppression,
    $g_{ut}$ and $g_{tt}$.
\end{itemize}
In a practical calculation it is cumbersome to implement the
renormalisation procedure in the described way, by providing
counterterms to $\epsilon^u_{jk}$ and $c,\ldots, b_{22}$.  Instead, it is
much easier to renormalise the $g_{jk}^{u,d}$.  If one renormalises all
$g_{jk}^{u,d}$ in the $\ov{\rm MS}$ scheme, one automatically complies with
SU(2)$\times$U(1) gauge symmetry. Therefore it is sufficient to choose
the  $g_{jk}^{d}$ in such a way that \eq{spu} is obeyed at tree level.
In our calculation we will only need
a counterterm to $g_{sb}$ (in addition to the usual QCD counterterms for
the SM parameters), if $m_s$ is set to zero.
If $m_s$ is kept non-zero, an additional counterterm to $g_{bs}$ is required.

In the three-spurion 2HDM the parameter $\tan\beta$ is
  well-defined, because unitary rotations of $(H_u,H_d)$ would lead
  to $\epsilon^d\neq 0$ at tree level and spoil the spurion expansion.
We are interested in phenomenologically interesting scenarios, in
  which the rare decays $B(B_s\to\mu^+\mu^-)$ or $b\to s\gamma$
  deviate from the SM predictions at a level probed in
current and forthcoming measurements. The corresponding decay
amplitudes involve a helicity flip and come with the Yukawa matrix $Y^d$, which
grows with  $\tan\beta$, so that we will consider the case that
$\tan\beta$ is large. An interesting feature of the three-spurion 2HDM
is that the above-mentioned amplitudes scale differently with
$\tan\beta$ than in the type-II model, due to new loop contributions
with $g_{ct}$.

Next we briefly discuss the leptonic Yukawa Lagrangian. The
extremely stringent experimental bounds on FCNC transitions like
$\ell_j \to \ell_k \gamma$ suggest that only one spurion is present in
the lepton sector, meaning that the lepton Yukawa couplings of the two
Higgs doublets are automatically diagonal in the mass eigenstate
basis.\footnote{We set neutrino Yukawa couplings to zero.}  For
simplicity, we choose the leptonic Lagrangian of type-II form (i.e.\ we
set $g^l = 0$) , with the familiar $\tan\beta$-enhanced
non-standard-Higgs couplings to charged leptons:
\begin{equation}
  \lag_{Y}^l =  \bar{l}_L \left[\frac{\hat{M}^l}{v}\tan\beta \right] l_R
  \frac{\phi^{0\prime} + \mathrm{i} A^0}{\sqrt2}
    + \bar{\nu}_L \left[\frac{\hat{M}^l}{v} \tan\beta \right] l_R H^+ +
  \hc
  \label{lyuk}
 \end{equation} 
The diagonal mass matrix of the charged leptons is denoted by $\hat{M}^l$
and the couplings of $\phi_{\rm SM}$ are omitted in \eq{lyuk}.

%- }}}
%- {{{ The decay $B_s \to \mu^+ \mu^-$:

\section{The decay $\mathbf{B_s \to \mu^+ \mu^-}$}\label{sec:bsmumu}

The typical momentum scale for $B_s$ decays is of order $M_{B_s}$ or smaller,
so that weak $B_s$ decays can be described by an effective theory in
which the heavy $W,Z$ bosons, the top quark, and the Higgs bosons of
the 2HDM are integrated out. The resulting $|\Delta B|=1$ Hamiltonian
$H_{\rm eff}$ describes the interactions mediated by these heavy particles in terms of
dimension-6 operators changing the beauty quantum number $B$ by one
unit. The piece of $H_{\rm eff}$  relevant for  $B_s \to \mu^+ \mu^-$ reads
\begin{equation}
H_{\mathrm{eff}} = N \sum_{i = A,S,P} \left(C_i Q_i + C_i' Q_i'
  \right) \,. \label{heff}
\end{equation}
The operators in \eq{heff} are 
\begin{equation}
\begin{alignedat}{3}
Q_A = \, & \left(\bar{b} \gamma_\mu P_L s \right) \left(\bar{\mu} \gamma^\mu \gamma_5 \mu \right) \,, && \qquad Q_A' = \, && \left(\bar{b} \gamma_\mu P_R s \right) \left(\bar{\mu} \gamma^\mu \gamma_5 \mu \right) \,, \\
Q_S = \, & \left(\bar{b} P_L s \right) \left(\bar{\mu} \mu \right) \,, && \qquad Q_S' = \, && \left(\bar{b} P_R s \right) \left(\bar{\mu} \mu \right) \,, \\
Q_P = \, & \left(\bar{b} P_L s \right) \left(\bar{\mu} \gamma_5 \mu \right) \,, &&\qquad Q_P' = \, && \left(\bar{b} P_R s \right) \left(\bar{\mu} \gamma_5 \mu \right) \,,
\end{alignedat}
\end{equation}
and are multiplied with their respective \emph{Wilson coefficients}
$C_A,\ldots,C_P'$ which contain the dependence on the large masses.
The normalisation factor in \eq{heff} is 
\begin{align}
 N& =\;  
    \frac{G_F^2 M_W^2 }{\pi^2} V_{ts} V_{tb}^* \,=\,
\frac{G_F \alpha_{\mathrm{em}}\left(\mu
    \right)}{\sqrt{2} \pi \sin^2 \theta_w} V_{ts} V_{tb}^*
    , \label{norm}
\end{align}
which complies with the conventions of Ref.~\cite{Hermann:2013kca}.  The
second ``$=$'' sign only holds to lowest order in the electroweak
interaction, while in higher orders the relation between the Fermi
constant $G_F$ and the electromagnetic coupling
$\alpha_{\mathrm{em}} =e^2/(4\pi)$, the weak mixing angle $\theta_w$ and
the $W$ mass $M_W$ is modified.  Electroweak corrections have been
calculated in Ref.~\cite{Bobeth:2013tba} and e.g.\ remove the
ambiguities related to the choice of the renormalisation scheme for these
parameters; in the second version for $N$ in \eq{norm} this issue also
includes the choice of the scale in the running 
$\alpha_{\mathrm{em}}$. In the following, we choose the first definition $N\propto
G_F^2$ in this paper, for which the electroweak corrections to the
SM contribution to $C_A$ are as small as $-2.4\%$
\cite{Bobeth:2013tba}.

We introduce the perturbative expansion
  of the $C_i$ as
\begin{equation}
C_i = C_i^{(0)} + \left(\frac{\alpha_s}{4\pi}\right)
C_i^{(1)} + \ldots \, ,
\end{equation}
where $ C_i^{(0)}$ denotes the leading order (LO),  arising  in the SM from
one-loop electroweak diagrams. $C_i^{(1)}$ comprises the
next-to-leading order (NLO) QCD corrections. 
In the SM only $C_A$ is relevant,
$C_A^\prime$ is suppressed w.r.t.\ $C_A$ by the ratio  $m_b m_s/M_W^2$ involving the strange and bottom quark masses $m_{s,b}$ and $C_{S,P}^{(\prime)}$ receive additional
suppression factors of $M_{B_s}^2/M_W^2$. 
The leading contributions
$C_i^{(0)}$ arise from one-loop electroweak diagrams at order $G_F^2$ in the
SM, hence the branching ratio is rather small.

The average time-integrated branching ratio is given
by~\cite{DeBruyn:2012wj,DeBruyn:2012wk}
\begin{align}\label{eq:timeavgbranchingratio}
  \overline{\mathcal{B}} \left(B_s \to \mu^+ \mu^-\right)  = \; 
  |N|^2 \frac{M_{B_s}^3 f_{B_s}^2}{32 \pi \Gamma_{H}^s} \beta
  \Big[ & \abs{r \left( C_A - C_A'\right) -
      u \left(C_P - C_P'\right)}^2 F_P  \;+ \nn
   & \abs{u \beta \left(C_S - C_S'\right)}^2 F_S \Big] \,,
\end{align}
with dimensionless quantities
\begin{equation}
r = \frac{2 m_\mu}{M_{B_s}} \,, \qquad \beta = \sqrt{1 - r^2} \,, \qquad u = \frac{M_{B_s}}{m_b + m_s} \,.
\end{equation}
Here, $\Gamma_H^s$ ($\Gamma_L^s$) denotes the decay width of the heavier
(lighter) $B_s$ mass eigenstate, and the factors $F_P$ and $F_S$ account for
the mixing of the $B_s-\bar{B}_s$ system, given by
\begin{align}
F_P = \, & 1 - \frac{\Gamma_L^s - \Gamma_H^s}{\Gamma_L^s} \sin^2 \left[ \arg\left(r \left(C_A' - C_A \right) - u \left(C_P' - C_P\right) \right) \right] \,, \nonumber\\
F_S = \, & 1 - \frac{\Gamma_L^s - \Gamma_H^s}{\Gamma_L^s} \cos^2 \left[ \arg \left(C_S' - C_S \right) \right] 
\end{align}
in the absence of significant CP-violating New Physics contributions to
the $B_s-\bar{B}_s$ mixing amplitude. In writing
\eq{eq:timeavgbranchingratio}, we have used the pseudoscalar decay
constant $f_{B_s}$ to rewrite the operator matrix elements as
\begin{align}\label{eq:hadrmatels}
\mel{0}{\bar{b} \gamma_\mu \gamma_5 s}{B_s (p)} = \, & \mathrm{i} p_\mu f_{B_s} \,, \nonumber\\
\mel{0}{\bar{b} \gamma_5 s}{B_s (p)} = \, & -\mathrm{i} f_{B_s} \frac{M_{B_s}^2}{m_b + m_s} \,,
\end{align}
where the second equation follows from the first one by use of the equations of motion.

The Higgs-mediated contributions in the SM can be neglected due to the tiny Yukawa couplings of external particles.
However, in the 2HDM with large Yukawa couplings of Higgs bosons to right-handed down-type quarks and leptons, Feynman diagrams with Higgs bosons are known to contribute significantly to $C_S$ and $C_P$, as will be discussed in the following.

%- }}}
%- {{{ Computational setup:

\section{Computational setup}\label{sec:setup}

In this section, we describe the chain of programs used to generate and
evaluate the corresponding Feynman diagrams at leading and
next-to-leading order.  We use \texttt{FeynRules} with the \texttt{Universal
  FeynRules Output} (\texttt{UFO})~\cite{Christensen:2009jx,Alloul:2013bka,Degrande:2011ua} to obtain Feynman
rules for the model.  The output is
processed by \texttt{tapir}~\cite{Gerlach:2022qnc} into a Lagrangian file, which is
used with \texttt{qgraf}~\cite{Nogueira:1991ex} in order to generate all
Feynman diagrams.  

We compute the one-loop diagrams for general electroweak gauge
parameters for the $W$ and $Z$ bosons and check that they drop out in
the final result. This is a welcome check for the conversion from
\texttt{UFO} to \texttt{qgraf} and \texttt{tapir}.  If diagrams with all
different quark flavours are explicitly calculated, we have at one-loop
level $<\mathcal{O}(100)$ Feynman diagrams in the SM,\footnote{From the
  technical perspective, it is convenient to split the electroweak gauge
  bosons into two different ``particles'', with different propagator
  denominators $\left(k^2 - M^2 \right)^{-1}$ and
  $\left(k^2 - \xi M^2 \right)^{-1}$, respectively, and treat them as
  different diagrams; hence the large number of
  diagrams in the 1-loop calculation.} and an additional
$\mathcal{O}(300)$ diagrams from contributions with at least one non-SM
Higgs boson, most of which are Higgs-penguin diagrams.
At two-loop order, we perform the calculation
in the Feynman gauge for the $W$ and $Z$ bosons, but we keep the gluon gauge parameter general and verify that it drops out of the final results.

The diagrams are then converted with the help of \texttt{tapir}
into \texttt{FORM}~\cite{Ruijl:2017dtg} code using Feynman rule definitions
that were also produced in the conversion from \texttt{UFO} to the
\texttt{qgraf} Lagrangian file.  {The individual expressions for the
  diagrams} are mapped onto integral families using
\texttt{exp}~\cite{Harlander:1998cmq,Seidensticker:1999bb} and a custom
\texttt{FORM} setup is used to perform the remaining computational
steps.  Since the Wilson coefficients are independent of the momenta of
the external particles, we set the latter to zero, so that only vacuum
integrals need to be computed. (An exception are diagrams with an
  FCNC self-energy in an external leg, which are calculated with
  on-shell $b$ quark and $m_b\neq 0$ before the subsequent limit
  $m_b\to 0$ is taken.)  We use a \texttt{FORM} implementation of the
  algorithm presented in Ref.~\cite{Davydychev:1992mt}, see
  Ref.~\cite{Salomon:2012iha}.

%- }}}
%- {{{ The decay $B_s \to \mu^+ \mu^-$ in the type-II Two-Higgs-Doublet Model

\section{The decay $\mathbf{B_s \to \mu^+ \mu^-}$ in the Two-Higgs-Doublet Model
  of type-II }\label{sec:decay2hdm}

\begin{figure}[t]
\centering
\includegraphics[width=\textwidth]{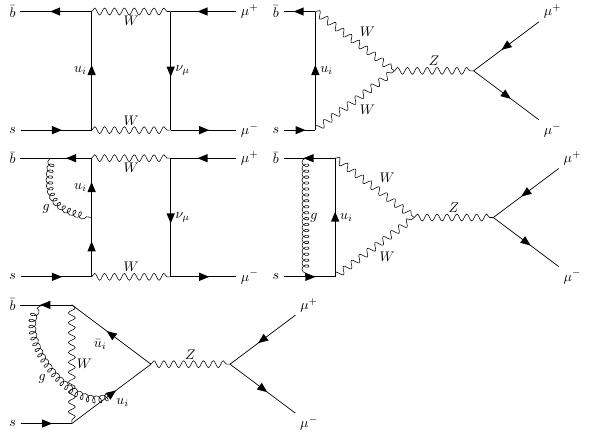}
\caption{\label{fig:SMdias}Sample diagrams contributing to $C_A$ at leading and next-to-leading
  order in the SM.}
~\\[-3mm]\hrule
\end{figure}

In the SM the leading-order result was obtained in Ref.~\cite{10.1143/PTP.65.297} and
next-to-leading order QCD corrections have been presented in
Refs.~\cite{BUCHALLA1993285,BUCHALLA1993225,Buchalla:1998ba,Misiak:1999yg}.
Higher-order QCD and electroweak corrections have been computed in
Refs.~\cite{Hermann:2013kca,Bobeth:2013tba,Bobeth:2013uxa,Beneke:2017vpq,Beneke:2019slt}.
Next-to-leading corrections in the type-II 2HDM have been calculated in
Refs.~\cite{Logan:2000iv,Bobeth:2001sq,Bobeth:2001jm,Hermann:2014lga}.
We have reproduced these results for the present paper, also as a
check of the automated setup.  The result can be expressed in terms of
dimensionless mass ratios of the top quark, $W$ boson and charged Higgs boson
masses,
\begin{align}
x_t &=\,  \frac{m_t^2}{M_W^2} \,,\qquad\qquad
r_H =\,  \frac{m_t^2}{M_{H^+}^2} \,.
\end{align}
In the SM, only the Wilson coefficient $C_A$ receives significant
contributions from diagrams such as the ones depicted in
\fig{fig:SMdias}.  The Wilson coefficients $C_A',C_S,\ldots,C_P'$
  are suppressed by powers of ratios of light (external) masses and $M_W$.  The
leading SM contribution is, moreover, independent of the Yukawa
couplings of all external particles, that is we can set
$m_b = m_s = m_\mu = 0$ for the contributions from $W$-box and
$Z$-penguin diagrams.  At leading and next-to-leading order they are
given by
\begin{align}
  C_A^{W+Z,(0)} = \,
  &   % \left[
    \frac{x_t \left(x_t - 4\right)}{8 \left(x_t - 1\right)} + \frac{3
    x_t^2 \log x_t}{8 \left(x_t - 1\right)^2} %\right]
    \,, \\
  C_A^{W+Z,(1)} = \,
  &% \left[
    \frac{ 2 x_t \left( 2 x_t^2 + 5x_t + 5 \right)}{3 \left(x_t -
    1\right)^2} - \frac{x_t \left(x_t^2 + 5 x_t + 2 \right) \log
    x_t}{ \left(x_t - 1\right)^3} - \frac{x_t \left(x_t^2 + 2\right)
    \mathrm{Li}_2 \left(1 - \frac{1}{x_t}\right)}{ \left(x_t -
    1\right)^2} % \right.
    \notag \\
  &+ %\left.
     \left( \frac{x_t \left(x_t^2 +
    x_t + 4\right)}{ \left(x_t - 1\right)^2} - \frac{6 x_t^2 \log
    x_t}{\left(x_t - 1\right)^3} \right) \log \frac{\mu^2}{m_t^2} %\right]
    \,, 
\end{align}
where the $Z$-penguin contributions include the flavour-changing
quark self-energy diagrams required to make the penguin
diagrams finite. 
Moreover, since charm and up quark masses can be neglected compared to
$M_W$ and $m_t$, after summation over internal up-type quarks all
contributions are proportional to the CKM factor $V_{ts}V_{tb}^*$
  (contained in the normalisation factor $N$ defined in \eq{norm}) due
to the unitarity of the CKM matrix.

\begin{figure}[t]
\centering
\includegraphics[width=\textwidth]{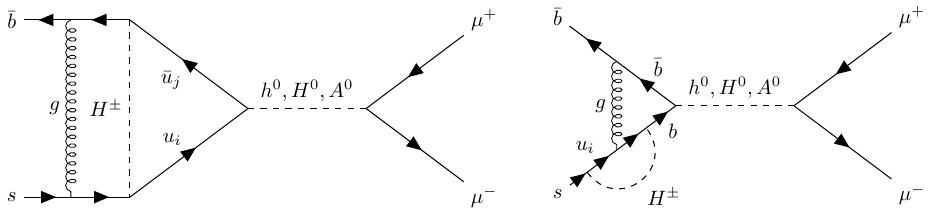}
\caption{Sample two-loop Feynman diagrams contributing to the Wilson
  coefficients $C_S^{(\prime)}$ ($H_{1,2}$) and $C_P^{(\prime)}$ ($A^0$). In the left diagram, only
  flavour-diagonal transitions $i = j$ are possible in the type-II 2HDM, while
  the more general 2HDM allows also for transitions with $i \neq j$,
  e.g. transitions from a charm quark into a top
  quark.}\label{fig:HiggspenFeynDia}
~\\[-3mm]\hrule
\end{figure}

In the 2HDM of type-II the leading new effects stem from
$\mathcal{O} \left(\tan^2 \beta \right)$ contributions to
$C_S^{(\prime)}$ and $C_P^{(\prime)}$, arising from penguin diagrams
with a neutral Higgs boson, see e.g.\ \fig{fig:HiggspenFeynDia}, and the
$W^+$-$H^+$ box diagram \cite{Logan:2000iv}. The Higgs-penguin
contributions to $C_S$ are given in Feynman gauge for the $W$-boson by
\begin{align}
  C_S^{h,(0)} = \, & - r_b \left[  \frac{r_H x_t \log r_H}{4 \left(r_H - 1\right) \left(r_H - x_t\right)} - \frac{r_H x_t \log x_t}{ 4 \left(x_t - 1 \right) \left(r_H - x_t \right)} \right] \,, \label{eq:CShpen1l} \\
  C_S^{h,(1)} = \, &  - r_b \left[ - \frac{8 r_H x_t}{3 \left(r_H - 1\right) \left(x_t - 1\right)} + \frac{2r_H x_t \left(3 r_H - 7 \right) \log r_H}{3 \left(r_H - 1\right)^2 \left(r_H - x_t \right)} - \frac{2r_H x_t \left(3x_t - 7 \right) \log x_t}{3 \left(x_t - 1\right)^2 \left(r_H - x_t\right)} \right. \notag \\
                   &- 2 \left. \log\left(\frac{\mu^2}{m_t^2}\right) \left( \frac{r_H x_t}{\left(r_H - 1\right) \left(x_t - 1\right)} + \frac{r_H x_t \log r_H}{\left(r_H - 1\right)^2 \left(r_H - x_t\right)} - \frac{r_H x_t \log x_t}{\left(x_t - 1\right)^2 \left(r_H - x_t\right)}\right) \right. \notag \\
                   &+ \left. \frac{2r_H x_t \mathrm{Li}_2 \left(1 - \frac{1}{r_H}\right)}{r_H - x_t} - \frac{2r_H x_t \mathrm{Li}_2 \left(1 - \frac{1}{x_t}\right)}{r_H - x_t} \right] \label{eq:CShpen2l} \,,
\end{align}
where
\begin{equation}
r_{q} = \frac{m_{\mu} m_q \, \tan^2\beta}{M_W^2} \,.
\end{equation}
The Wilson coefficients for the right-handed operators $C_S'^{h,(i)}$ can be obtained by the replacement $r_b \to r_s$ in \eqsand{eq:CShpen1l}{eq:CShpen2l}.
Further contributions include terms of order
  $m_b m_s m_\mu^2
    \tan^4 \beta / M_W^4$ and $m_t^2 m_\mu^2 /
    M_W^4 $ entering the Wilson coefficients $C_A^\prime$ and
  $C_A$, respectively, as well as $\mathcal{O} \left(m_b m_s \tan^2
    \beta / M_W^2 \right)$ (in $C_A'$) and $\mathcal{O} \left(m_t^2 \cot^2
    \beta / M_W^2 \right)$ (in $C_A$) terms arising from $Z$-penguin
  diagrams with a charged Higgs boson.
Box diagrams with a single charged Higgs boson also contribute to
$C_{S,P}$ ($C_{S,P}'$), with Wilson coefficients proportional to $m_b$
($m_s$).
We do not explicitly list these contributions here; they can be found in Refs.~\cite{Bobeth:2001sq,Bobeth:2001jm,Hermann:2014lga}.
With the exception of the doubly muon-mass suppressed $H^+$--$H^-$ box contributions to $C_A$, we include all of these additional terms in our analysis.
The feature that $C_{S}$ and $C_{P}$ are proportional to  $m_b$
while their primed counterparts are proportional to $m_s$
holds beyond the type-II 2HDM in our more general 2HDM with three
  spurions because of $\epsilon^d = P_d \left(\epsilon^u, Y^u, Y^d
\right) Y^d$, entailing factors of $m_{d_j}$ in Yukawa couplings of
$d_{jR}$. A rather remarkable feature of the type-II 2HDM is the fact that the leading
terms in $\tan\beta$ depend only on $\tan\beta$ and the charged-Higgs boson
mass $M_{H^+}$, that is they are independent of the parameters of the neutral
Higgs sector~\cite{Logan:2000iv}.
In the type-II 2HDM, the leading
$\tan^2 \beta$ contributions to the pseudoscalar and scalar Wilson
coefficients satisfy the rather simple relation
\begin{equation}\label{eq:typeIIrelationCSCP}
C_S = C_P \,, \qquad C_S' = -C_P' \,.
\end{equation}

\begin{figure}[t]
\centering
\includegraphics[width=\textwidth]{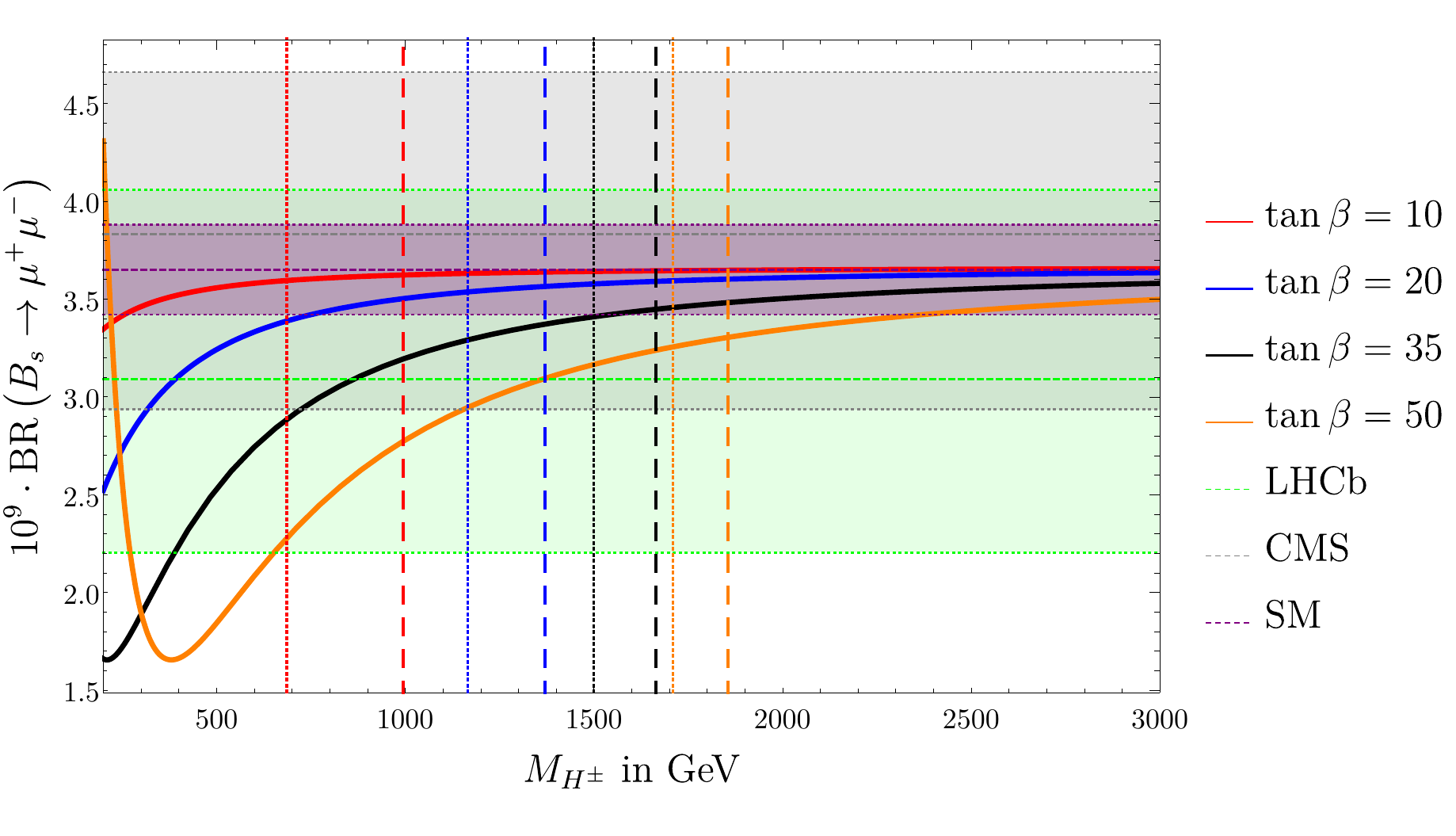}
\caption{\label{fig::BR_2hdm}Branching ratio
  $\mathcal{B}\left(B_s \to \mu^+ \mu^- \right)$ in the type-II
  Two-Higgs-Doublet model at next-to-leading order in QCD, for different
  values of $\tan \beta$. All running parameters are evaluated at the
  scale $\mu = \overline{m}_t \left(\overline{m}_t\right)$.
  The green dashed and dotted lines denote the central value and the
  experimental $2 \sigma$ error of
      the LHCb measurement \cite{LHCb:2021vsc,LHCb:2021awg},
      respectively, the grey band shows the corresponding information for
      the CMS result \cite{CMS:2022dbz}. The purple dashed and dotted
    bands indicate the SM prediction with uncertainties, presented in
    Ref.~\cite{Bobeth:2013uxa}. Note that the SM prediction in the
    purple band was obtained including also
    $\mathcal{O} \left(\alpha_s^2\right)$ and
    $\mathcal{O} \left(\alpha_{\mathrm{em}} \right)$
    contributions and corresponds to $|V_{cb}|=0.0424\pm 0.0009 $.
    The perturbativity of the scalar 2HDM potential constrains
      $\abs{M_{A^0}^2 - M_{H^\pm}^2} \leq \abs{\lambda_4 - \lambda_5}
      v^2 \lesssim 8
      v^2$~\cite{Cacchio:2016qyh,Chowdhury:2015yja,Casalbuoni:1986hy,Maalampi:1991fb,Kanemura:1993hm,Akeroyd:2000wc,Ginzburg:2005dt,Horejsi:2005da,Haber:2010bw,Grinstein:2015rtl,Durand:1992wb,Maher:1993vj,Durand:1993vn,Nierste:1995zx},
      which can be converted into a lower bound on $M_{H^\pm}$ by use of
      the experimentally excluded area of the $(M_{A^0},\tan\beta)$ plane
      such as the ones shown in \fig{fig:ATLASHiggsExclusion} used by us. The
      vertical dashed lines indicate these lower limits on $M_{H^\pm}$ for each
      value of $\tan\beta$, whereas the vertical dotted lines show the
      same limits obtained with the less stringent constraint
      $\abs{\lambda_4 - \lambda_5} \leq 8 \pi$. The plot shows that
        $B_s \to \mu^+ \mu^-$ will only give contraints on the type-II
        2HDM competitive with the Higgs searches, once the experimental
        precision on $\mathcal{B}\left(B_s \to \mu^+ \mu^- \right)$
        substantially improves.} 
  ~\\[-3mm]\hrule
\end{figure}
In Fig.~\ref{fig::BR_2hdm} we show the branching ratio
$\mathcal{B}\left(B_s \to \mu^+ \mu^- \right)$ in the type-II 2HDM as a
function of the charged Higgs boson mass. The horizontal green,
grey, and violet bands correspond to the experimental measurements
of LHCb~\cite{Zyla:2020zbs} and CMS~\cite{CMS:2022dbz}
as well as the theory
prediction~\cite{Bobeth:2013uxa}, respectively, including
2$\sigma$
% 1~sigma
uncertainties for the experimental values (cf.~Table.~\ref{tab:numinput}). 
The  theory prediction of Ref.~\cite{Bobeth:2013uxa} uses
  $|V_{cb}|=0.0424\pm 0.0009 $, which is close to today's value inferred
from inclusive $b\to c\ell\nu$ decays. If one uses $|V_{cb}|=0.03936\pm
0.00068$ from exclusive $B$ decays \cite{HFLAV:2019otj} the central
value of the theory prediction for $10^9 \mathcal{B}\left(B_s \to \mu^+
  \mu^- \right)$ drops from $3.65$ to $3.15$.

The coloured lines are predictions from the 2HDM for different values of
$\tan\beta$. It is interesting to note that for $\tan\beta \lesssim 25$ low
values of $M_{H^\pm}$ are required to reproduce the central value of the
LHCb measurement.  Of course, in the limit $M_{H^\pm}\to\infty$ all 2HDM
curves approach the SM prediction. Note that in the type-II model there
is a $\tan\beta$-independent 95\% C.L. lower bound on $M_{H^\pm}$ in the
range 570-800 GeV from $B\to X_s \gamma$~\cite{Misiak:2017bgg}.  This
bound can be easily weakened with our model's additional couplings
discussed in the following section.

Recently it has been pointed out that LHC data still permit
$M_{H^\pm}\leq 400\,$GeV with couplings compatible with solutions of the
$b\to c \tau \nu$ flavour anomalies
{\cite{Iguro:2022uzz,Blanke:2022pjy}}. Charged-Higgs explanations of
the latter have been found viable in
Refs.~\cite{Blanke:2018yud,Blanke:2019qrx} and are invigorated by recent
LHCb data on $B\to D^{(*)} \tau \nu$ \cite{LHCbSem}, see
Refs.~\cite{Iguro:2022yzr,Fedele:2022iib}.  {It should be clearly
  stated that this solution to the $b\to c \tau \nu$ puzzle is not
  realised in the type-II model, in which for instance
  $\mathcal{B}(b\to c \tau \nu)$ is suppressed rather than enhanced over
  $\mathcal{B}(b\to c \mu \nu)$ as preferred by data. Also} while
charged Higgs searches at the LHC are compatible with the quoted $H^\pm$
masses \cite{ATLAS:2018gfm}, the lower bound on $M_{H^\pm}$ inferred
from the data on $gg/b\bar b \to A^0[\to\tau^+\tau^-]$ searches is
larger than 1 TeV \cite{ATLAS:2022yvl} in perturbed versions of the
type-II 2HDM. {Also in our model we cannot substantially weaken this
  bound; to this end one must modify}  the $A^0$ coupling to $\tau$'s
{by deviating from the type-II form in \eq{lyuk} and permitting the third
  lepton generation to couple to the other Higgs doublet}.

%- }}}
%- {{{ Additional contributions in a model with flavour-changing neutral Yukawa couplings:

  \section{Additional contributions in a model with flavour-changing
    neutral Yukawa couplings}\label{sec:newcontributions}

\begin{figure}[t]
\centering
\includegraphics[width=.6\textwidth]{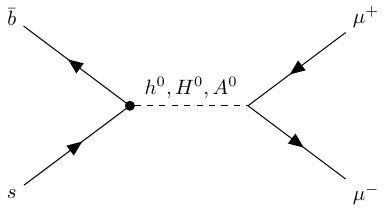}
\caption{Tree-level diagrams with flavour-changing neutral Higgs
  couplings. The $b-s-H$ coupling is denoted by a dot on the vertex.}\label{fig:gsbdias}
\end{figure}

\begin{figure}[t]
\centering
\includegraphics[width=0.6\textwidth]{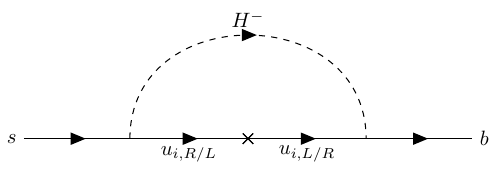}
\caption{The loop-induced change of flavours $s \to b$ via a quark
  self-energy diagram. The helicity flip denoted by the cross
  entails a factor of $\tan\beta$ and linearity in
  $\hat{g}^u$.}\label{fig:selfenergy}
~\\[-3mm]\hrule
\end{figure}

  In a model with a Yukawa Lagrangian given by \eq{eq:yuklag} there are
  additional tree-level contributions
  $\bar{b} s \to h^0 (H^0, A^0) \to \mu^+ \mu^-$ of order $\tan\beta$. A
  sample Feynman diagram is shown in \fig{fig:gsbdias}. At loop-level
  there are ${\cal O}(\tan^3 \beta)$ terms due to diagrams in which
    the neutral Higgs boson couples to the $b$ line. At LO these are
    diagrams with a FCNC self-energy and we also refer to them as
    self-energy diagrams at NLO, even if a gluon connects the FCNC loop
    with the $s$ quark as in the diagram on the right in
    \fig{fig:HiggspenFeynDia}.  These
  $\mathcal{O}\left(\tan^3\beta\right)$ terms occur because a
  $\tan \beta$-enhanced coupling in the self-energy diagrams is not
  cancelled by a factor $\cot\beta$ in the second charged-Higgs
  coupling, which is a distinguishing feature compared to the
  type-II model.  If these self-energy diagrams involve a helicity flip
  of the internal fermion line, see \fig{fig:selfenergy}, they come with
  a factor of $\tan \beta$ and are linear in the flavour-changing Yukawa
  matrix $g^u$, which enters the result through the
    charged-Higgs coupling in the fourth term of \eq{eq:yuklag}.  In
    fact, the dominant dependence on $g_{ct}$ stems from this source and
    not from diagrams in which a neutral Higgs boson couples to charm
    and top in a vertex diagram, which have a factor of $\tan\beta$
    less.  Note that contributions from diagrams without helicity flip
  are either already included in the type-II model or quadratic in
  $g^u$ (and without the factor $\tan\beta$), and we will
  consequently neglect the latter in what follows.

\subsection{Pseudoscalar Wilson coefficient $C_P$}
At tree level the Wilson coefficient $C_P$ originating from 
the diagram in \fig{fig:gsbdias} with a virtual $A^0$ boson is given by
\begin{equation}
  C_{P}^{(0),\mathrm{tree}}
    = - \frac{\pi^2}{G_F^2 M_W^2 V_{tb}^* V_{ts}} \frac{m_{\mu}\,\tan\beta }{2 v M_{A^0}^2} g_{sb}^* =
    -\frac{g_{sb}^*}{g_{bs}}
    C_{P}^{\prime (0)} \,.
  \label{eq:CP_tree}
\end{equation}
Since there is no loop-suppression, even small values of $g_{bs}$ and
$g_{sb}^*$ can have significant impact on the branching ratio,
but the spurion expansion in \eq{spu} naturally leads to (parametrically
  suppressed) small couplings.
For convenience, let us define the linear combinations
  $g_{bs}^{\pm} \equiv g_{bs} \pm g_{sb}^*$.

At one-loop order only the self-energy diagrams contain an enhancement
  factor $\tan^3\beta$ whereas the vertex contributions contain at most a quadratic
  term. The one-loop self-energy contributions to $C_P$ and $C_P^{\prime}$ are ultra-violet
  divergent. The corresponding counterterm is generated from
\eq{eq:CP_tree} and is of the form
\begin{eqnarray}
  g_{bs}^{+,0} &=& g_{bs}^+ + \delta_{A^0,bs}^{+\,(0)} +
                   \left(\frac{\alpha_s}{4\pi}\right) \delta_{A^0,bs}^{+\,(1)} + \ldots  \,.
\end{eqnarray}
In the $\overline{\rm MS}$ renormalisation scheme the one-loop
contribution is given by
\begin{align}\label{eq:CPA0CTLO}
    \delta_{A^0,bs}^{+\,(0)} \, =\; - \frac{1}{\epsilon} \frac{\sqrt{2} G_F
      m_t \tan^2\beta}{8 \pi^2}
                            & \left[ \;\left(  g_{ut}^* V_{us} + 
                            g_{ct}^* V_{cs} +  g_{tt}^* V_{ts}
                            \right) m_b V_{tb}^* \right. \notag \\
                               & \left. +
                            \left( g_{ut}V_{ub}^* + g_{ct}
                                 V_{cb}^*  + g_{tt}  V_{tb}^*
                                 \right) m_s V_{ts} \right]    \,.
\end{align}
Defining analogously
\begin{eqnarray}
  g_{bs}^{-,0} &=& g_{bs}^- + \delta_{A^0,bs}^{-\,(0)} +
                   \left(\frac{\alpha_s}{4\pi}\right)
                   \delta_{A^0,bs}^{-\,(1)} + \ldots  \, ,
\end{eqnarray}
the counterterms $ \delta_{A^0,bs}^{-\,(0)}$ and
$\delta_{A^0,bs}^{-\,(1)}$ are found from
$ \delta_{A^0,bs}^{+\,(0)}$ and $\delta_{A^0,bs}^{+\,(1)}$ in
\eqsand{eq:CPA0CTLO}{eq:CPA0CTNLO} below by the replacement $m_b\to -m_b$.
  Thus the terms with $m_b$ renormalise the Wilson coefficient
$C_P \propto  g_{bs}^+- g_{bs}^-$,
while the ones with $m_s$ renormalise the Wilson coefficient $C_P^\prime
\propto  -\left(g_{bs}^++ g_{bs}^-\right)$.
The renormalised finite pseudoscalar Wilson coefficients read
\begin{equation}\label{eq:CPtanbeta3LO}
\begin{aligned}
  C_P^{(0)} = \, & C_{P}^{(0),\mathrm{tree}} + \, 
  \tilde{N} \,\frac{ m_b}{M_{A^0}^2}\,
  \left(g_{ut}^* \frac{V_{us}}{V_{ts}} + g_{ct}^* \frac{V_{cs}}{V_{ts}} + g_{tt}^*  \right)  \, \tilde{C}_P^{(0)} \,, \\
  C_P^{\prime(0)} = \, & C_{P}^{\prime(0),\mathrm{tree}} \,  - \tilde{N} \,
  \frac{ m_s}{M_{A^0}^2}  \left(g_{ut} \frac{V_{ub}^*}{V_{tb}^*} + g_{ct} \frac{V_{cb}^*}{V_{tb}^*} + g_{tt} \right)  \, \tilde{C}_P^{(0)} \,,
\end{aligned}
\end{equation}
with normalisation factor
\begin{equation}
  \tilde{N} = \frac{m_t m_{\mu} \tan^3 \beta}{G_F \sqrt{2}
    M_W^2 v } \,=\,
  \frac{g m_t m_{\mu} \tan^3 \beta}{2 G_F 
    M_W^3 }, 
\end{equation}
the weak coupling constant $g$, and
\begin{equation}
  \tilde{C}_P^{(0)} = -\frac{1}{8} \left[1 +
    \log \frac{\mu^2}{m_t^2}  - \frac{\log r_H}{r_H - 1} \right] \,.\label{eq:tcp0}
\end{equation}
{Our results in \eqsto{eq:CP_tree}{eq:tcp0} confirm the result in
  Eq.~(3.16) of Ref.~\cite{Crivellin:2019dun}.}  At NLO in QCD (i.e.\
two-loop order), the counterterm required to obtain a finite result is
given by
\begin{align}\label{eq:CPA0CTNLO}
      \delta_{A^0,sb}^{(1)} =\; -\frac{G_F m_t \tan^2 \beta}{\sqrt{2} \pi^2}
          &\left[ \; \left(  g_{ut}^* V_{us} +  g_{ct}^* V_{cs} + 
                               g_{tt}^* V_{ts} \right) m_b V_{tb}^* \right. \notag \\
                             & \left. + \left(  g_{ut} V_{ub}^* +  g_{ct} V_{cb}^*
                               +  g_{tt}  V_{tb}^* \right) m_s
                               V_{ts} \right]\times \left[ \frac{1}{\epsilon^2} +
                           \frac{4}{3\epsilon} \right] \,.
\end{align}
Note that in addition to the top quark mass $m_t$ also 
the flavour-changing Yukawa couplings are minimally renormalised.
The two-loop contribution to the Wilson coefficients reads 
\begin{equation}\label{eq:CPtanbeta3NLO}
\begin{aligned}
  C_P^{(1)} = \, & \;\;\;
  \tilde{N} \, \frac{ m_b}{M_{A^0}^2}\,
  \left(g_{ut}^* \frac{V_{us}}{V_{ts}} + g_{ct}^* \frac{V_{cs}}{V_{ts}} + g_{tt}^*  \right)  \,
   \tilde{C}_P^{(1)} \,, \\
    C_P'^{(1)} = \, & - \tilde{N} \,
     \frac{ m_s}{M_{A^0}^2}  \left(g_{ut} \frac{V_{ub}^*}{V_{tb}^*} + g_{ct} \frac{V_{cb}^*}{V_{tb}^*} + g_{tt} \right)  
   \, \tilde{C}_P^{(1)} \,,
\end{aligned}
\end{equation}
with
\begin{align}
\tilde{C}_P^{(1)} =& 
-\frac{4 (r_H-2)}{3 (r_H-1)}
+\frac{(3 r_H-7)
   \log (r_H)}{3 (r_H-1)^2} 
+ \left(\frac{7-4 r_H}{3 (r_H-1)}
   -\frac{\log (r_H)}{ (r_H-1)^2}\right) \log \left(\frac{\mu ^2}{m_t^2}\right) \notag \\ 
   &-\frac{1}{2} \log
   ^2\left(\frac{\mu ^2}{m_t^2}\right)
   + \text{Li}_2\left(1-\frac{1}{r_H}\right) \,.
\end{align}
Next we discuss the dependence of our result on the renormalisation
  scale $\mu$ at which the 2HDM result is matched to the effective
  $|\Delta B|=1$ Hamiltonian and the size of higher-order corrections.
  In the case $m_t \sim  M_{H^+}$ the choice $\mu={\cal O}
  (m_t,M_{H^+})$ leads to the absence of large logarithms and the
  variation of $\mu$ between $m_t$ and $ M_{H^+}$ does not constitute
  a relevant source of theoretical uncertainty. Thus we are left to the
  phenomenologically interesting case $ M_{H^+} \gg m_t$.  
The $\log\mu^2$ terms in the Wilson coefficients have two different
origins, divergent contributions involving Yukawa couplings or the QCD
coupling, respectively. $C_P^{(0)}$ stems from a loop with 
Yukawa couplings and $C_P^{(0)}$ involves no large logarithm for the choice
$\mu \sim M_{H^+}$. To verify this expand \eq{eq:tcp0} as
\begin{equation}
  \tilde{C}_P^{(0)} = -\frac{1}{8} \left( 1 +
    \log \frac{\mu^2}{M_{H^+}^2} \right) \left( 1 + {\cal O} \lt(
    r_H\rt)  \right)\,. 
  \label{eq:tcp0exp}
\end{equation}
This feature is generic for all $\log\mu^2$ terms stemming from loops
with Yukawa couplings, because heavy particles like $H^+$ do not
contribute to the renormalisation group functions (i.e.\ $\beta$
functions and anomalous dimensions) for $\mu < M_{H^+}$ since heavy
particles are integrated out at scales of the order of their masses.
The same loop diagrams yielding $ \tilde{C}_P^{(0)}$ also
determine the piece of the $\beta$ functions of $g_{sb}$ proportional to the Yukawa
couplings of top and bottom quarks.  The running of $g_{sb}(\mu)$
from this source is compensated by the explicit logarithm in
\eq{eq:tcp0exp} and the remaining Yukawa-$\mu$ dependence is a tiny
two-loop effect. 

\begin{figure}
\centering
\includegraphics[width=.8\textwidth]{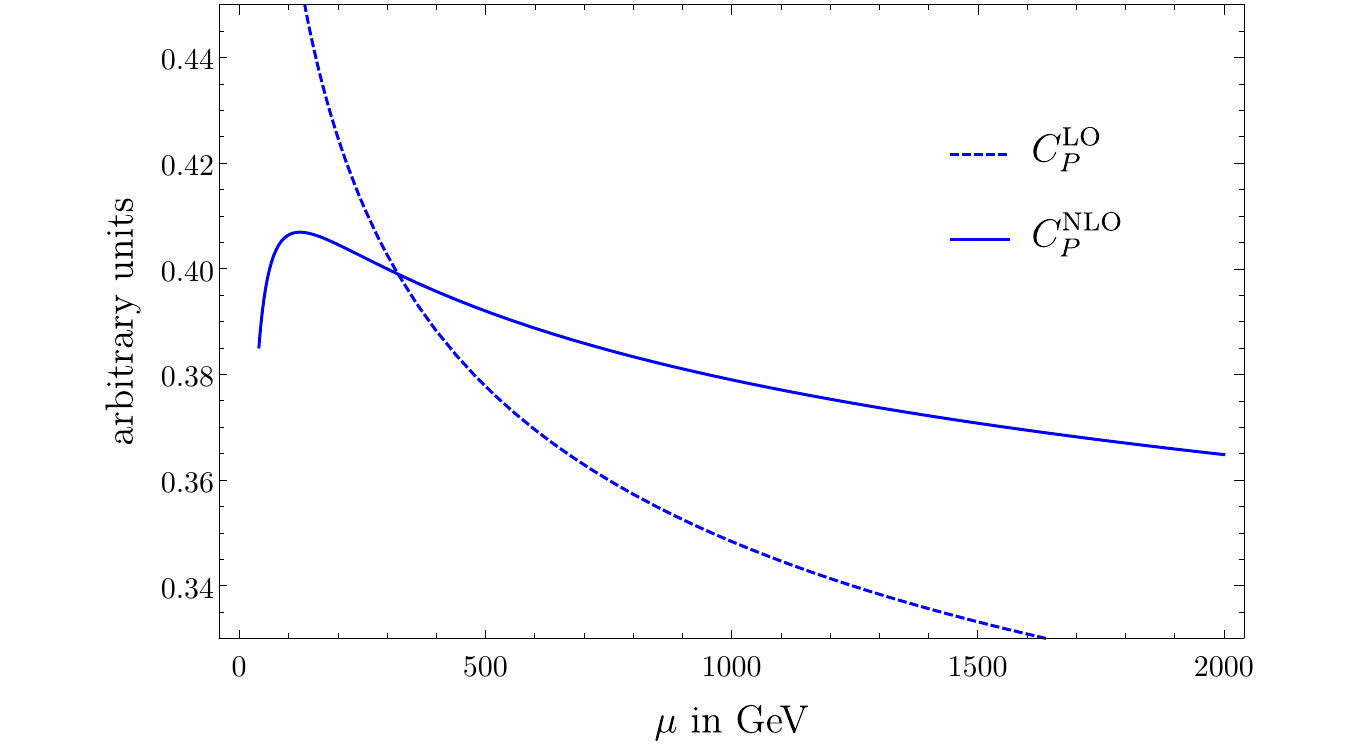}
\caption{QCD scale dependence of the Wilson coefficient $C_P$ at
    leading and next-to-leading order in $\alpha_s$.
In this plot we have used $M_{H^+} = \SI{1.5}{\tera\eV}$ and $g_{ct}
\left(\overline{m}_t \right) = 1$
and have further set $V_{ts}/V_{cs} \to 0$.
All $\mu$-independent prefactors have been fixed.
$C_P(\mu)$ depends logarithmically on the masses $m_t$ and $M_{H^+}$, so
  that every choice of $\mu$ in the interval $[m_t,M_{H^+}]$ seems
  justified. The plot shows, however, that choosing $\mu \sim  M_{H^+}$
  in $C_P^{(0)}(\mu)$ would badly underestimate the NLO result, while $\mu
  \sim  m_t$ would  sizably overestimate it. $C_P^{(0)}(m_t)$
  exceeds $C_P^{(0)}( M_{H^+})$ by 24\%, while the corresponding value
  for $C_P^{(1)}(\mu)$ is 9\%.\label{fig:mudependence}}
~\\[-3mm]\hrule
\end{figure}

The situation is different with the $\mu$ dependence stemming from gluon
loops and related to the familiar QCD running of couplings and quark
masses. The QCD running of $g_{sb}$ is trivial, since the combination
$g_{sb}(\mu) / m_b (\mu)$, which enters $C_P \cdot \expval{Q_P}$, is independent of $\mu$
(see \eq{eq:hadrmatels}). To study the $\mu$ dependence of our two-loop
result $C_P^{(1)}$, we keep the $\log\mu^2$ term stemming from the
Yukawa interaction (i.e. the analogue of the logarithm in
\eq{eq:tcp0exp}) fixed and vary $\mu$ otherwise.  $C_P^{(0)}$ depends on
$\mu$ implicitly through the $\mu$-dependence of $m_t$ and $g_{it}$ and
this dependence should be compensated by the explicit $\log\mu^2$ terms
in $C_P^{(1)}$, reducing the $\mu$-dependence to the three-loop
(one-loop Yukawa correction and NNLO QCD) level.\footnote{The scale dependence of the light quark mass $m_b \left(\mu\right)$ in $C_P$ is a relic of our choice of definition of the effective operators, and will be cancelled by the corresponding running of the hadronic matrix element, see \eq{eq:hadrmatels}. Thus, we do not consider the running of $m_b$ in the following.}
In \fig{fig:mudependence} we illustrate the QCD scale dependence of the
$\mathcal{O}\left(\alpha_s^0 \right)$ and
$\mathcal{O} \left(\alpha_s \right)$ Wilson coefficient $C_P$ for a
particular choice of $M_{H^+}$ and
$g_{ct} \equiv g_{ct} \left(\bar{m}_t\right)$.  All parameters that are
not running in QCD have been fixed in this figure and we have
neglected the contributions from $g_{ut}$ and $g_{tt}$, which have the
same running as $g_{ct}$, such that the only running parameters are in
$m_t g_{ct}^* \tilde{C}_P^{(0,1)}$.  The figure shows a
significant improvement of the QCD scale dependence with the inclusion
of next-to-leading order QCD corrections. The LO result does not
permit a reliable prediction of $C_P$, while 
$C_P^{\rm NLO}$ merely changes by $\pm 5\%$ around its central value
when $\mu$ is varied between $m_t$ and $M_{H^+}$.

From \fig{fig:mudependence} it is clear that our calculated QCD
  corrections are needed for a reliable prediction. Next we discuss
  uncalculated higher-order 
  corrections involving Yukawa couplings, obtained by dressing the LO
  diagrams with an additional Higgs boson. The large ${\cal O}(1)$
  couplings are the coupling of the SM-like Higgs boson $h^0$ to the top quark
  and  the $A^0$, $H^0$ couplings to the bottom quark.\footnote{While a
    priori the couplings of $\phi_{\rm new}$ to up-type quarks could be
  $\geq {\cal O}(1)$, we will see in Sec.~\ref{sec:phenopart} that in
the phenomenologically interesting parameter region they are smaller.}
The former
  contributions are already present in the SM, contained in the
  electroweak corrections of Ref.~\cite{Bobeth:2013tba}. Since they are very
  small in the SM, they will constitute an even smaller correction to the extra
  diagrams of the 2HDM. The dominant contribution from an extra loop
  with $A^0$ or $H^0$ is expected from diagrams in which both ends of
  the additional Higgs line are connected to a $b$ line. Contrary to the
  QCD case, all $A^0$, $H^0$ diagrams
  involve momenta which are far off-shell, because $M_{A^0,H^0}$ is much
  larger than $m_b$. Thus the additional loop will give a correction in
  the few-\% region to the leading 2HDM term, which is numerically constrained to
  the range between SM prediction and experimental value and thereby
  constitute a small correction to a small LO 2HDM term.

\subsection{Scalar Wilson coefficient $C_S$}

The scalar Wilson coefficients receives contributions from both neutral CP-even Higgs mass eigenstates $h^0$ and $H^0$.
At tree level, the diagrams with $h^0$ and $H^0$ give rise to the Wilson coefficients
\begin{equation}
    C_{S}^{(0),\mathrm{tree}} = - \frac{\pi^2}{G_F^2 M_W^2 V_{tb}^* V_{ts}} \frac{m_\mu\,\tan\beta }{2 v} R_M g_{sb}^*
    = \frac{g_{sb}^*}{g_{bs}} C_{S}^{\prime(0)} \,,
\end{equation}
where 
\begin{equation}
R_M = \frac{M_{h^0}^2 \sin^2 \left(\beta - \alpha\right) + M_{H^0}^2 \cos^2 \left(\beta - \alpha\right)}{M_{h^0}^2 M_{H^0}^2}
\end{equation}
contains the dependence on the neutral Higgs-boson masses.
The counterterms required to cure the divergences at one and two loops can be obtained from \eqsand{eq:CPA0CTLO}{eq:CPA0CTNLO} 
analogously, i.e.\ the combination $g_{bs}^+ - g_{bs}^-$ renormalises $C_S$, while the combination $g_{bs}^+ + g_{bs}^-$ renormalises $C_S^\prime$.
The renormalised Wilson coefficients $C_S$ and $C_S'$ are related to the pseudoscalar ones by
\begin{equation}
C_S^{(i)} = R_M \, M_{A^0}^2 \, C_P^{(i)} \,, \qquad C_S'^{(i)} = -R_M \,M_{A^0}^2 \, C_P'^{(i)} \,.
\end{equation}
Note that there are no QCD corrections to $C_{S,P}^{(\prime),\mathrm{tree}}$ since they cancel in the matching calculation.

\section{Phenomenology}\label{sec:phenopart}
In this section we discuss the possible size  of $\mathcal{B}\left(B_s \to \mu^+ \mu^-
  \right)$ in our 2HDM under the constraint that other $b\to s$
  processes comply with the data.
We include the tree-level contributions from diagrams with
flavour-changing down-type couplings (cf.~\fig{fig:gsbdias}), as well as
the SM results and the leading quadratic $\tan\beta$ contributions in the type-II
2HDM to which the additional diagrams of order $\tan^3 \beta$ discussed in the previous section add as corrections.

In a generic 2HDM the tree-level couplings $g_{bs}^{\pm}$ will
drastically increase the branching ratio for $B_s \to \mu^+ \mu^-$
due to the missing loop suppression. In our model the spurion
  expansion suppresses $g_{bs}^{\pm}$ in a controlled way, but
  still permits large enough contributions to get phenomenologically
  interesting effects in  $B_s \to \mu^+ \mu^-$:
 Even rather small up-type couplings $g_{ct}$ significantly modify the
 Wilson coefficients $C_P^{(\prime)}$ and $C_S^{(\prime)}$, as they
 feature a CKM factor $V_{cs}$ instead of $V_{ts}$.  In the following,
 we will restrict ourselves to the experimentally favoured scenario
 \cite{ATLAS:2019nkf,CMS:2018uag} of aligned Higgs doublets, in
 which $\sin\left(\beta - \alpha\right) \approx 1$.  For the numerical
 analysis, we will set $\sin\left(\beta - \alpha\right) = 1$, and
 therefore have $R_M = M_{H^0}^{-2}$.  In this case, the Higgs mass
 eigenstates $h^0$ and $H^0$ coincide with $\phi^0$ and
 $-\phi^{0\prime}$, respectively, and only the latter possesses
 non-SM-like couplings to fermions.  This reduces the number of relevant
 non-Yukawa-type parameters of the extended Higgs sector to four, namely
 $\tan\beta$, $M_{H^+}$, $M_{A^0}$ and
 $M_{\phi^{0\prime}} \equiv M_{H^0}$.

\subsection[Constraints from $b \to s \gamma$ decays]{Constraints from $\mathbf{b \to s \gamma}$ decays}
\begin{figure}
\centering
\includegraphics[width=0.5\textwidth]{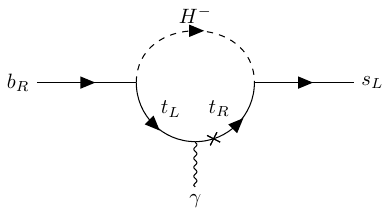}
\caption{Sample Feynman diagram for the rare decay $b \to s \gamma$ at one loop in the 2HDM. The cross indicates a chirality flip $t_L \to t_R$.}\label{fig:btosgammaFeynmandia}
~\\[-3mm]\hrule
\end{figure}
An important constraint on the magnitude of flavour-changing Yukawa couplings in the up-type quark sector arises from the inclusive rare decays $B \to X_s \gamma$.
This process is mediated at the quark level by $b \to s \gamma$
through a top
quark loop with a charged $W$ boson in the Standard Model and receives
additional contributions through charged-Higgs boson diagrams in the
2HDM, see \fig{fig:btosgammaFeynmandia}.
In order to eliminate the dependence on $V_{cb}\simeq -V_{ts}$,
one traditionally works with the quantity
\begin{equation}
R_\gamma \equiv \frac{\mathcal{B} \left(b \to s \gamma \right) +
  \mathcal{B} \left(b \to d \gamma \right)}{\mathcal{B} \left(b \to c l
    \nu \right)} .
\end{equation}
Note that in the SM and the type-II 2HDM
the branching ratio of $b \to d \gamma$ is much smaller than
that of $b \to s \gamma$. 

In the type-II 2HDM, the contribution of the charged Higgs diagrams to
the decay rate is positive, moving the theoretical prediction for the
branching ratio away from the experimental value averaged
in \cite{Misiak:2017bgg} to

\begin{equation}\label{eq:Rgammaexp}
R_{\gamma,\mathrm{exp.}} = \left(3.22 \pm 0.15 \right) \cdot 10^{-3} \,,
\end{equation}
where a lower cutoff of $E_0 \geq \SI{1.6}{\giga\eV}$ has been put on
the photon energy.  The practical independence of $R_\gamma$ of
$\tan \beta$ in the largest part of the parameter space in the type-II
2HDM allowed for the extraction of a lower limit on the mass of the
charged Higgs boson in \cite{Misiak:2017bgg}, see
\autoref{sec:decay2hdm}.  In the general 2HDM, where, contrary to
  the type-II case, the factor $\tan \beta$ from the $btH^-$ vertex
is not cancelled by a factor $\cot \beta$ from the $stH^-$ vertex,
the quantity $R_\gamma$ depends on $\tan\beta$.\footnote{The absence
  of the $\cot \beta$ suppression of the $stH^-$ vertex is also a
  feature of the aligned 2HDM \cite{Jung:2010ik,Jung:2010ab,Jung:2012vu}.}
The combination of the
$\bar t_L b_RH^+$ and $\bar s_L t_R H^-$ Yukawa couplings arising in the
process is then given by
\begin{equation}\label{gsteff0}
\left(\frac{m_b \tan\beta\, V_{tb}}{v}\right) \left(\frac{m_t
    \,V_{ts}^*}{v \tan\beta} + V_{us}^* g_{ut} + V_{cs}^*\, g_{ct} + V_{ts}^*\, g_{tt}
\right) =
    \frac{m_b m_t \, V_{tb} V_{ts}^*}{v^2} \left(1 + g_{st}^{\mathrm{eff}} \right) \,,
\end{equation}
where we have defined the short-hand notation
\begin{equation}
  g_{st}^{\mathrm{eff}} \equiv \frac{v \, \tan\beta}{m_t}
  \left(g_{ut} \frac{V_{us}^*}{V_{ts}^*}
      + g_{ct}
    \frac{V_{cs}^*}{V_{ts}^*} + g_{tt} \right) \,. \label{gsteff}
\end{equation}
Note that $g_{st}^{\mathrm{eff}}$ carries a factor of $\tan\beta$.
The abbreviation $g_{st}^{\mathrm{eff}}$ denotes the size of
  the additional $\bar{s}_L t_R H^+$ coupling of our three-spurion 2HDM (terms involving $g^u_{ij}$) in units of
  the same coupling in the type-II 2HDM (denoted by ``$1$'' in
  \eq{gsteff0}).  Thus, the limiting case of the type-II 2HDM is given
  by $g_{st}^{\mathrm{eff}} = 0$, while our effective $H^\pm$
    coupling vanishes for $g_{st}^{\mathrm{eff}} = -1$, i.e.\
    $g_{st}^{\mathrm{eff}} = -1$ must be chosen to recover the SM result
    for $R_\gamma$. We further note that
  $g_{st}^{\mathrm{eff}}$ is dominated by $g_{ct}$ due to the large
  ratio $\abs{V_{cs}/V_{ts}}$, and neglecting subleading
  CKM-matrix elements (amounting to set to zero $g_{tt}$ and $g_{ut}$, which is
   instead constrained by $b\to d$ processes) allows to convert constraints on
  $g_{st}^{\mathrm{eff}}$ into bounds on $g_{ct}$ in the following
  discussion.

\begin{figure}
\centering
\includegraphics[width=\textwidth]{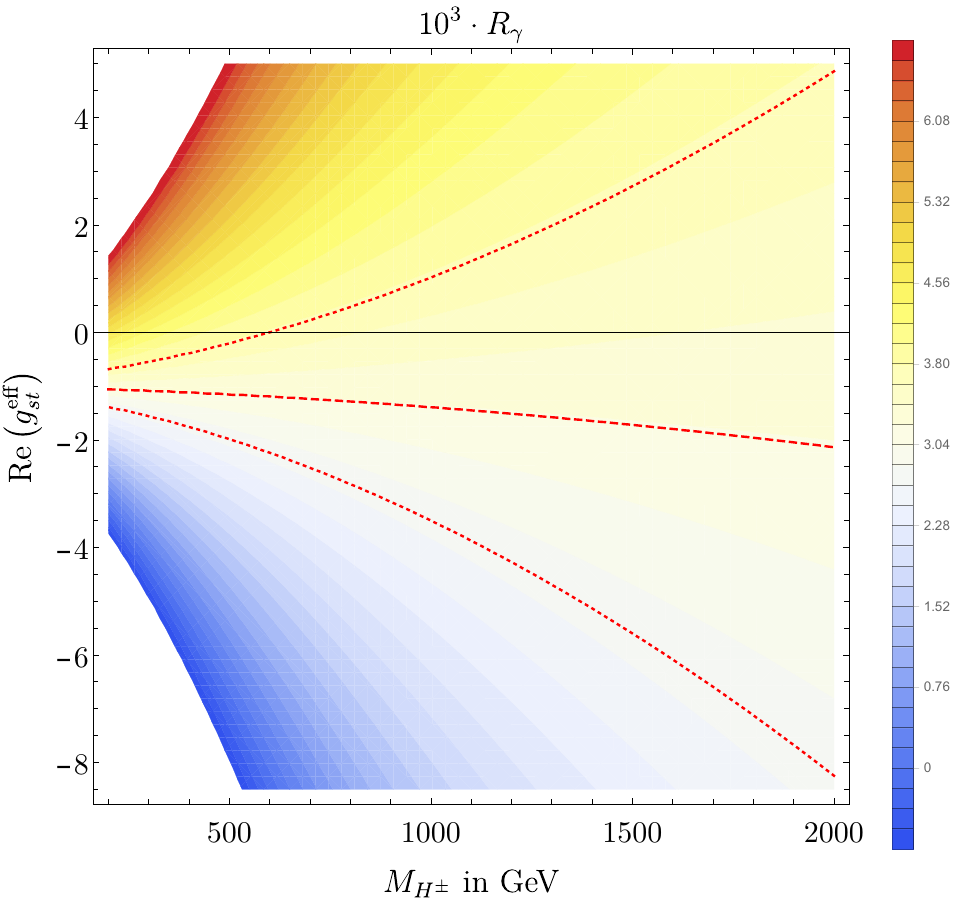}
\caption{The ratio $R_\gamma$ as a function of the charged Higgs boson mass $M_{H^\pm}$ and $g_{st}^{\mathrm{eff}}$. The dashed lines correspond to the experimental central value and the $\pm 2 \sigma$ intervals, to which we have also added the theoretical uncertainties.}\label{fig:btosgammaRgammacontourwiththeouncertainty}
~\\[-3mm]\hrule
\end{figure}

In order to constrain our new  flavour-changing couplings, we use the
results from \cite{Hermann:2012fc} and \cite{Misiak:2017bgg} with
this trivial change of the $\bar s_Lt_RH^-$ Yukawa coupling.
For large enough values of $M_{H^\pm}$, the central value of $R_{\gamma}$ is approximately given (at $\mu = \overline{m}_t \left(\overline{m}_t \right)$) by 
\begin{align}\label{eq:Rgammaapprox}
R_{\gamma} \approx 10^{-4} \cdot \Bigg\{\left.33.10\right|_{\mathrm{SM}}
  + & \left(1 +\,
  \mathrm{Re}\, g_{st}^{\mathrm{eff}} \right) \left[ r_H \left(-48.93 -
      47.60 \log r_H - 0.99 \left(\log r_H \right)^2
      \right. \right.  \notag \\ &
\left. \left. \left. -0.15 \left(\log r_H \right)^3 + 4.71 \,
                                          \mathrm{Li}_2 \left(1 -
                                          \frac{1}{r_H}\right) \right)
                                          \right. \right. \notag \\ &
\left. \left. + r_H^2 \left(-53.82 - 98.18 \log r_H + 4.79 \,
                                                                      \mathrm{Li}_2 \left(1 - \frac{1}{r_H}\right) \right) \right. \right. \notag \\ &
\left.  + r_H^3 \left(-56.04 - 150.43 \log r_H + 3.17 \, \mathrm{Li}_2 \left(1 - \frac{1}{r_H}\right) \right)\right] \Bigg\} \,,
\end{align}
which agrees with the exact result within $1 \, \%$ in the complete
subdomain of
$ \left(M_{H^\pm}, \mathrm{Re}\,g_{st}^{\mathrm{eff}} \right) \in
\left[\SI{500}{\giga\eV},\infty \right) \times \left[-5,5\right]$ in
  which $R_\gamma$ lies within the band allowed by the
  experimental and theoretical uncertainties (see below). The
  approximate formula in \eq{eq:Rgammaapprox} only includes the
  interference of the new-physics contribution with the SM result and
  neglects the squared new-physics contribution. In deriving \eq{eq:Rgammaapprox} we have used Eq.~(10) of Ref.~\cite{Misiak:2015xwa}.  We adopt the estimate
of the theoretical uncertainty of about $6.73\,\%$ given in
\cite{Misiak:2017bgg}, consisting of individual uncertainties of $5\,\%$
(non-perturbative), $1.5\,\%$ (parametric), $3\,\%$ (higher-order), and
$3\,\%$ (interpolation in the charm quark mass $m_c$).  In
\fig{fig:btosgammaRgammacontourwiththeouncertainty}, we illustrate the
ratio $R_\gamma$ in the $\left(M_{H^\pm}, \mathrm{Re}\,g_{st}^{\mathrm{eff}}\right)$
plane. The thick dashed line corresponds to the central experimental
value in \eq{eq:Rgammaexp} and
$R_{\gamma,\mathrm{exp}} \pm \Delta_{\mathrm{exp} + \mathrm{th}}$, where
$\Delta_{\mathrm{exp} + \mathrm{th}} = 2 \sigma_{\mathrm{exp}} +
\delta_{\mathrm{th}}$ is the sum of the experimental $2\sigma$
uncertainty intervals and the theoretical uncertainty of $6.73\,\%$.
For small values of $M_{H^\pm}$, the allowed range of
$g_{st}^{\mathrm{eff}}$ is tightly constrained around
$g_{st}^{\mathrm{eff}} = -1$, due to the closeness of the Standard Model
prediction and the experimental central value.  At larger $M_{H^\pm}$, a
significantly wider range of flavour-changing Yukawa couplings is
allowed.  From $R_\gamma$ we derive $M_{H^\pm}$-dependent upper and
lower bounds on $\mathrm{Re}\left(g_{st}^{\mathrm{eff}} \right)$ which
we will use in order to constrain the real part of the flavour-changing
up-type couplings appearing in $B_s \to \mu^+ \mu^-$.

\begin{figure}
\centering
\includegraphics[width=\textwidth]{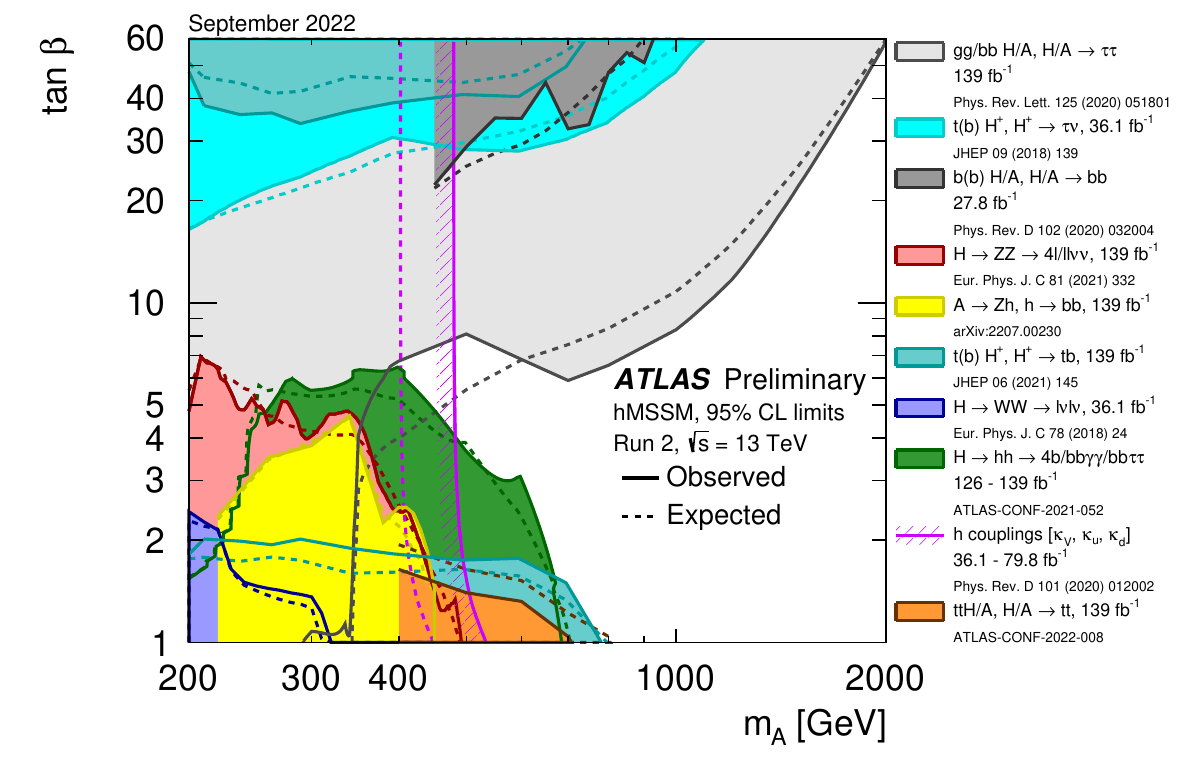}
\caption{Exclusion limits in the $M_{A^0} - \tan\beta$ plane from
  different searches performed by the ATLAS experiment; the plot is taken from
  Ref.~\cite{ATLAS:2022yvl}. The shown limits are for a different variant of
  the type-II 2HDM than ours, but qualitatively apply as well to our
  model.}\label{fig:ATLASHiggsExclusion}
~\\[-3mm]\hrule
\end{figure}

\subsection{Higgs searches}
Searches for heavy Higgs bosons at
the LHC put powerful lower limits on the masses of new
  Higgs particles, but these depend on the Yukawa sector of the
  considered 2HDM. For the case of the type-II model (or
  ramifications of it) these searches already exclude a significant
portion of the $M_{A^0} - \tan\beta$ plane at present.  In
\fig{fig:ATLASHiggsExclusion}, we show a recent collection of
exclusion limits obtained through various different searches by the
ATLAS experiment~\cite{ATLAS:2022yvl} in a modification of the
  type-II model.  These limits imply that large
values of $\tan\beta$ can only be realised if at the same time the
additional Higgs bosons are quite heavy,
$M_{A^0} \gtrsim \SI{1.5}{\tera\eV}$.

It should be noted that for heavy $M_{A^0}$ these bounds also
approximately apply for $M_{H^+}$ and $M_{H^0}$, since the masses become
degenerate in the heavy-Higgs limit (for a thorough analysis of the
  allowed mass splittings see \cite{Cacchio:2016qyh}; see also the vertical dashed lines in \fig{fig::BR_2hdm}).  Note that the
  bounds from neutral Higgs searches are more constraining than those
  for $H^+$ searches \cite{ATLAS:2018gfm}. The most stringent
  constraints are from final states with $\tau$'s and also apply to our
  lepton Yukawa Lagrangian in \eq{lyuk} which we have chosen of type-II.
  (However, the choice of the $\tau$ Yukawa couplings is of no relevance
  for the phenomenology of the $b\to s$ FCNC processes discussed in this
  paper.) The presence of the additional couplings $g_{jk}^u$ will
    increase some branching ratios at the expense of those of decays
    into $\tau$'s, so that these bounds can be somewhat weakened, but
    the general trend remains valid. Our scenarios discussed below
      comply with the ATLAS bounds.

\subsection[$B_s - \bar B_s$ mixing]{$\mathbf{B_s - \bar B_s}$ mixing}
The effective $\bar s_L b_R A^{0}$ and $\bar s_L b_R H^0$ vertices
  are the sum of the tree-level couplings $\propto g_{sb}$ and the loop
  contributions involving vertex and self-energy diagrams
  $\propto g_{ut} V_{tb} V_{us}^* + g_{ct} V_{tb} V_{cs}^* + g_{tt} V_{tb} V_{ts}^*$.  
  The coefficients $C_P$ and
  $C_S$ are proportional to this effective vertex, which is therefore
  the relevant quantity constrained by
  $\mathcal{B}(B_s\to \mu^+\mu^-) $.  Now the $B_s - \bar B_s$ mixing
  amplitude depends quadratically on this effective vertex, while both
  quantities decrease quadratically with $M_{A^0}$ and $M_{H^0}$, so that
  the quantities give complementary information on the parameter space.
  $B_s - \bar B_s$ mixing is mediated in the Standard Model by the
$\abs{\Delta B}=2$ operator
$Q_{VLL} = \left(\bar{b}_L \gamma_\mu s_L \right) \left(\bar{b}_L
  \gamma^\mu s_L \right)$ and in the 2HDM by three additional effective
operators
\begin{equation}
\begin{alignedat}{1}
Q_{SLL} = \, & \left(\bar{b}_R s_L \right) \left(\bar{b}_R s_L \right) \,, \\
Q_{SLR} = \, & \left(\bar{b}_R s_L \right) \left(\bar{b}_L s_R \right) \,, \\
Q_{SRR} = \, & \left(\bar{b}_L s_R \right) \left(\bar{b}_L s_R \right) \,,
\end{alignedat}
\end{equation}
The Wilson coefficient of $Q_{SLL}$ only involves the effective
  vertex governing $B_s \to \mu^+ \mu^-$ and the tree-level propagators
  of $A^0$ and $H^0$, while the coefficients of the other
  operators also involve the corresponding chirality-flipped effective
  $\bar s_R b_L A^{0}$ and $\bar s_R b_L H^0$ vertices.  The
coefficient $C_{SLL}$ ($C_{SRR}$) is proportional to $m_b^2$ ($m_s^2$),
while the coefficient $C_{SLR}$ is proportional to $m_b m_s$; we
therefore drop the operator $Q_{SRR}$ in our analysis.  However,
considering the ATLAS constraints implying large masses $M_{A^0,H^0}$,
$C_{SLL}\propto 1/M_{A^0}^2-1/M_{H^0}^2$ is heavily suppressed due
to the small splitting of Higgs masses in the large-mass
limit, 
and $C_{SLR}$ becomes relevant. This feature
  results from the fact that $Q_{SLL}$ violates hypercharge by two units
  of the Higgs hypercharge and the coefficient $C_{SLL}$ is therefore
  suppressed by a factor of $v^2/M_{A^0}^2$ compared to $C_{SLR}$ 
  multiplying  $Q_{SLR}$ which conserves hypercharge and SU(2).
  As a consequence, $C_{SLR}$ is more important than $C_{SLL}$. This
  feature has been widely studied in the context of the effective 2HDM
  emerging from integrating out superpartners in the MSSM
  \cite{Isidori:2001fv,Buras:2002wq,Buras:2002vd,Gorbahn:2009pp}.   

In the following we correlate  $\mathcal{B}(B_s\to \mu^+\mu^-) $
  with  $\mathcal{B}(B\to X_s \gamma) $ and the $B_s-\bar B_s$
  oscillation frequency $\Delta M_{B_s}$, which is proportional to the
  magnitude of the  $B_s-\bar B_s$ mixing amplitude.
  Only the effective  $\bar s_L b_R A^{0}$ and $\bar s_L b_R H^0$
  vertices are physical, by e.g.\ changing the renormalisation condition
  for $g_{sb}$ we can shift pieces between $g_{sb}$ and
  the renormlised loop. 
For simplicity, we set the tree-value of $g_{sb}$ to zero, i.e.\ consider the
  case that this coupling is only generated radiatively. The only
  non-trivial Yukawa structure entering the considered observables is then
$g_{st}^{\mathrm{eff}}$.

\begin{figure}
\centering
\begin{subfigure}[t]{.45\textwidth}
\centering
\includegraphics[width=0.97\textwidth]{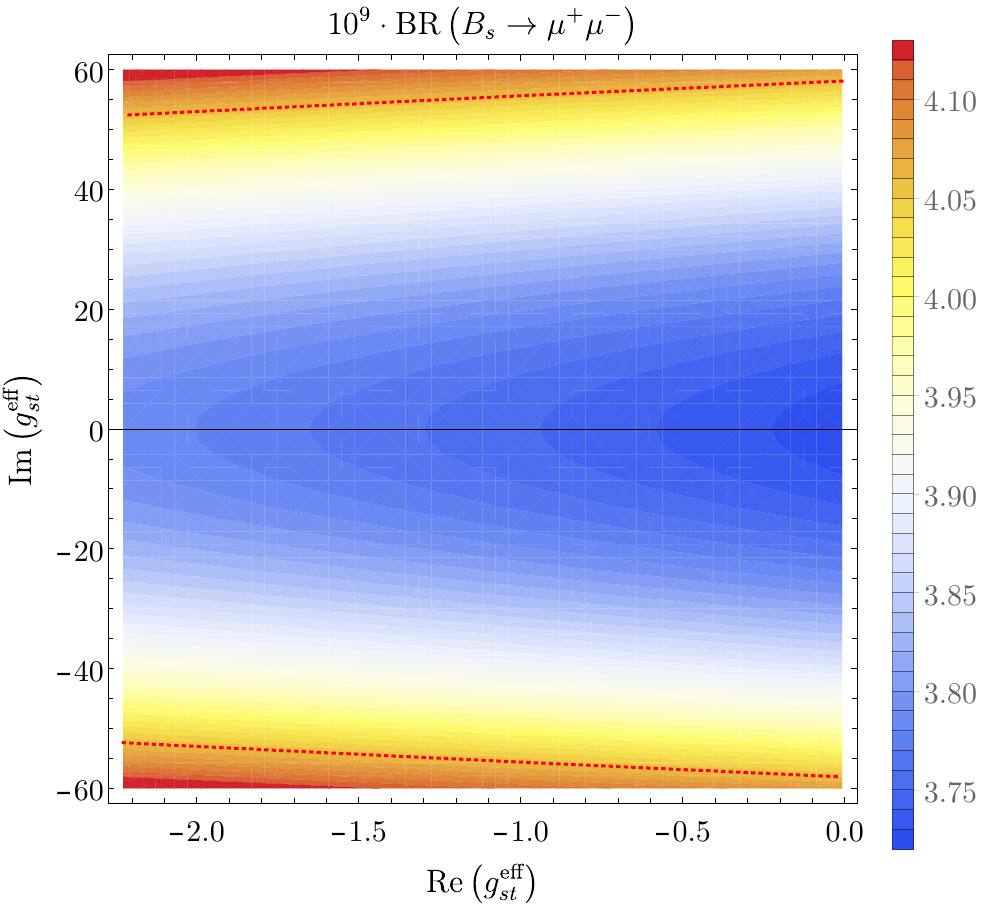}
\caption{$M_{H^\pm} = M_{H^0} = M_{A^0} = \SI{600}{\giga\eV}$, $\tan\beta = 7$.}
\end{subfigure}
\begin{subfigure}[t]{.45\textwidth}
\centering
\includegraphics[width=0.97\textwidth]{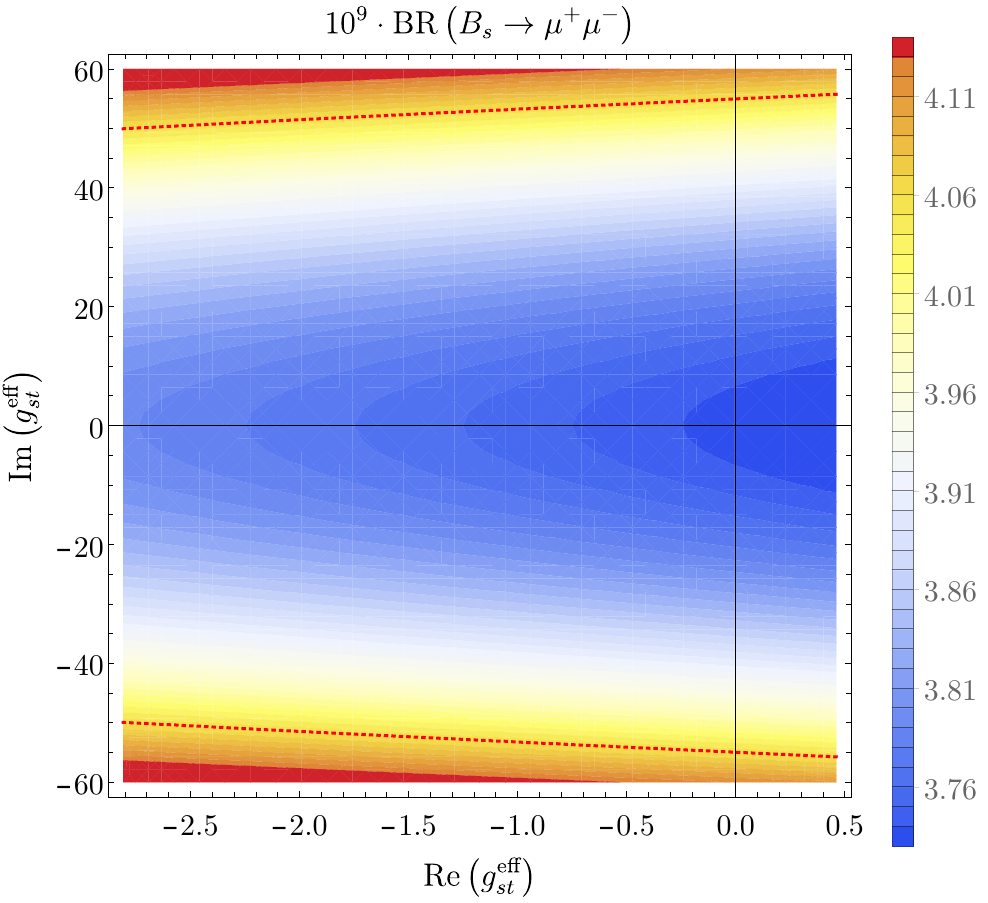}
\caption{$M_{H^\pm} = M_{A^0} = \SI{800}{\giga\eV}$, $M_{H^0} = \SI{600}{\giga\eV}$, $\tan\beta = 7$.}
\end{subfigure}
\begin{subfigure}[t]{.45\textwidth}
\centering
\includegraphics[width=0.97\textwidth]{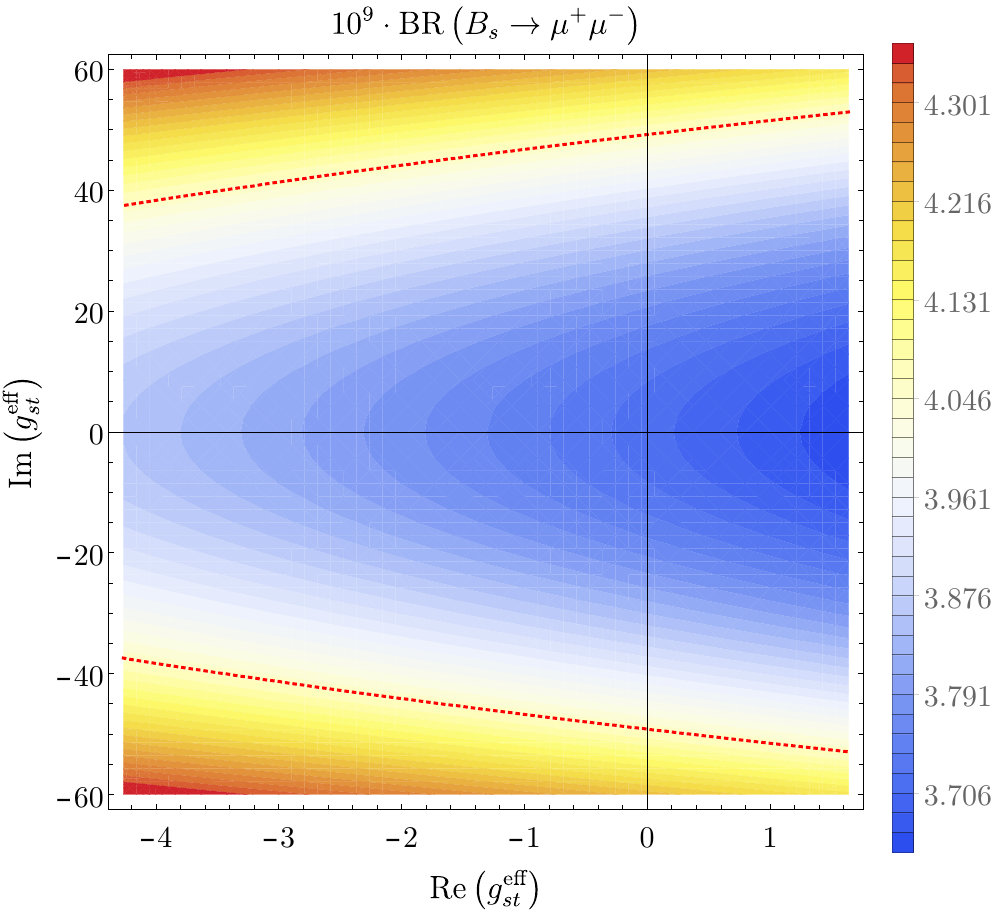}
\caption{$M_{H^\pm} = M_{H^0} = M_{A^0} = \SI{1200}{\giga\eV}$, $\tan\beta = 12$.}
\end{subfigure}
\begin{subfigure}[t]{.45\textwidth}
\centering
\includegraphics[width=0.97\textwidth]{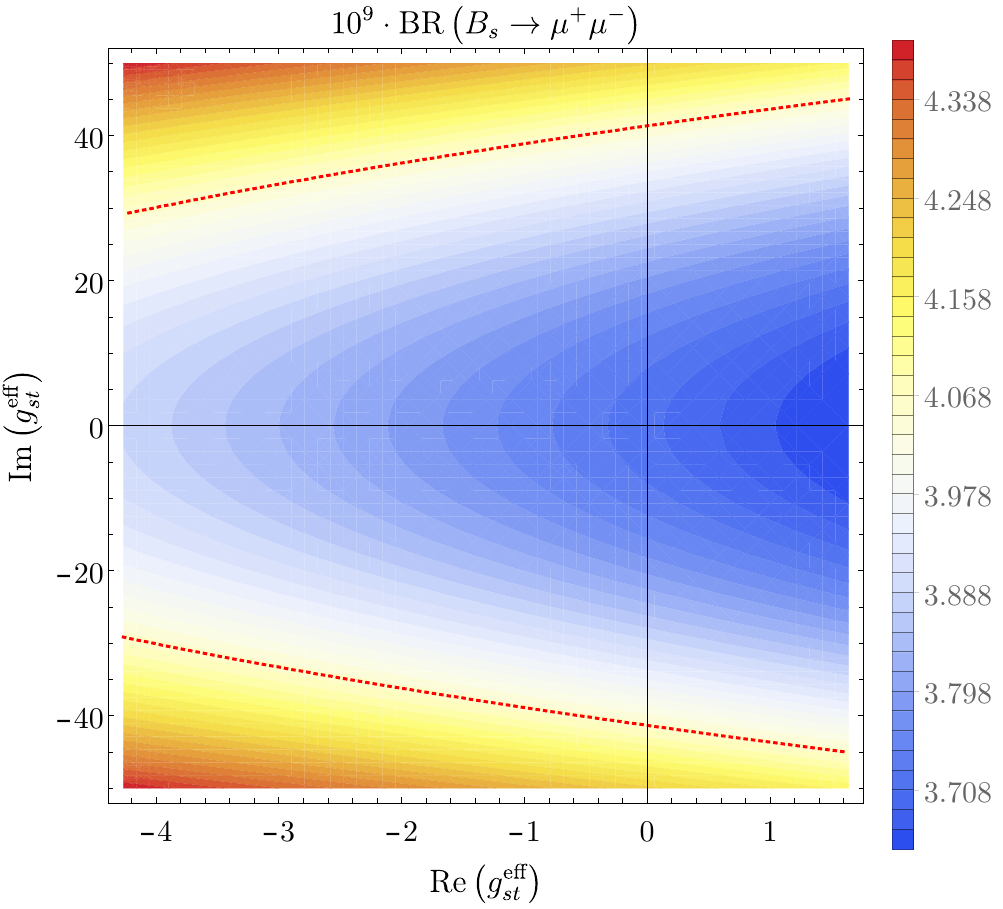}
\caption{$M_{H^\pm} = \SI{1200}{\giga\eV}$, $M_{H^0} = M_{A^0} = \SI{1100}{\giga\eV}$, $\tan\beta = 12$.}
\end{subfigure}
\begin{subfigure}[t]{.45\textwidth}
\centering
\includegraphics[width=0.97\textwidth]{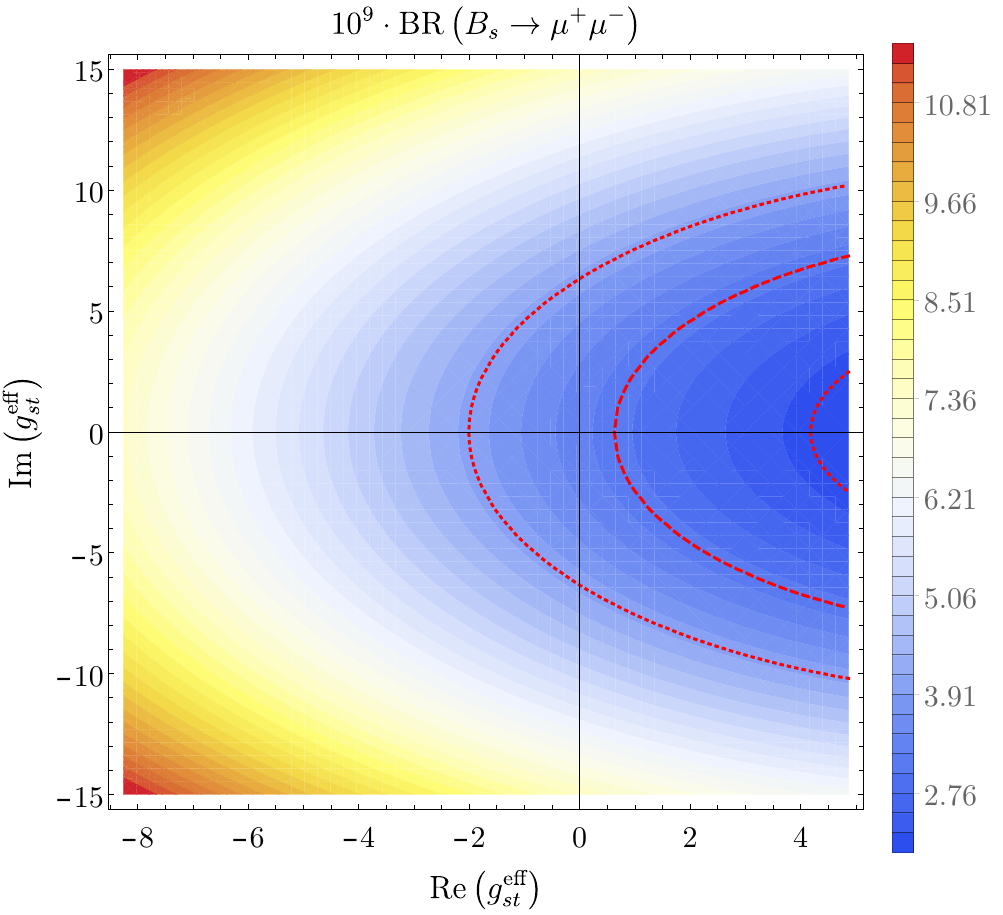}
\caption{$M_{H^\pm} = M_{H^0} = M_{A^0} = \SI{2000}{\giga\eV}$, $\tan\beta = 60$.}
\end{subfigure}
\begin{subfigure}[t]{.45\textwidth}
\centering
\includegraphics[width=0.97\textwidth]{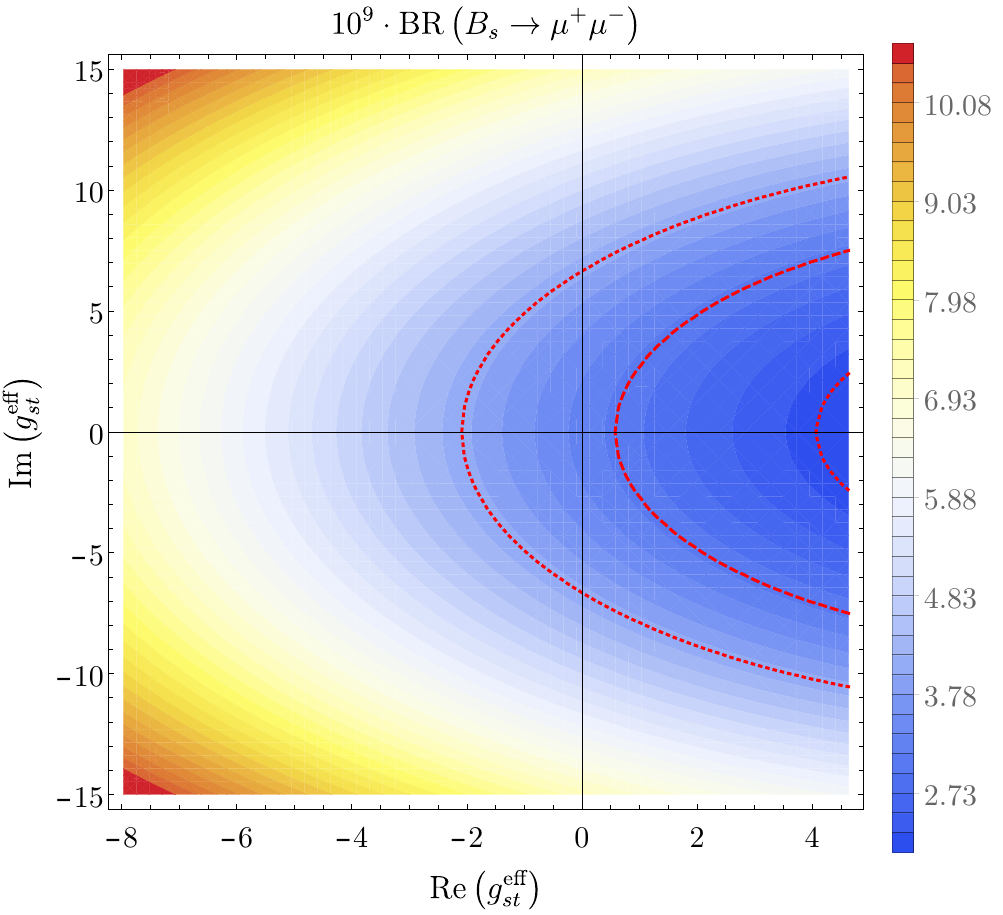}
\caption{$M_{H^\pm} = \SI{1950}{\giga\eV}$, $M_{H^0} = \SI{2050}{\giga\eV}$, $M_{A^0} = \SI{2000}{\giga\eV}$, $\tan\beta = 60$.}
\end{subfigure}
\caption{Branching ratio of $B_s\to \mu^+\mu^-$ for different values of
    Higgs masses and $\tan\beta$. All quantities are evaluated at
  $\mu = \overline{m}_t \left(\overline{m}_t\right)$. The red dashed and
  dotted lines indicate the experimental central value and the $2\sigma$
  uncertainties of LHCb branching ratio
  measurement~\cite{LHCb:2021vsc,LHCb:2021awg}, respectively.
  The constraint of $\mathcal{B}(B\to X_s\gamma)$ on $|\mathrm{Im}\,
  g_{st}^{\mathrm{eff}}|$ is not shown.\label{fig:phenoplots1}} 
~\\[-3mm]\hrule
\end{figure}

\begin{figure}
\centering
\begin{subfigure}[t]{.45\textwidth}
\centering
\includegraphics[width=\textwidth]{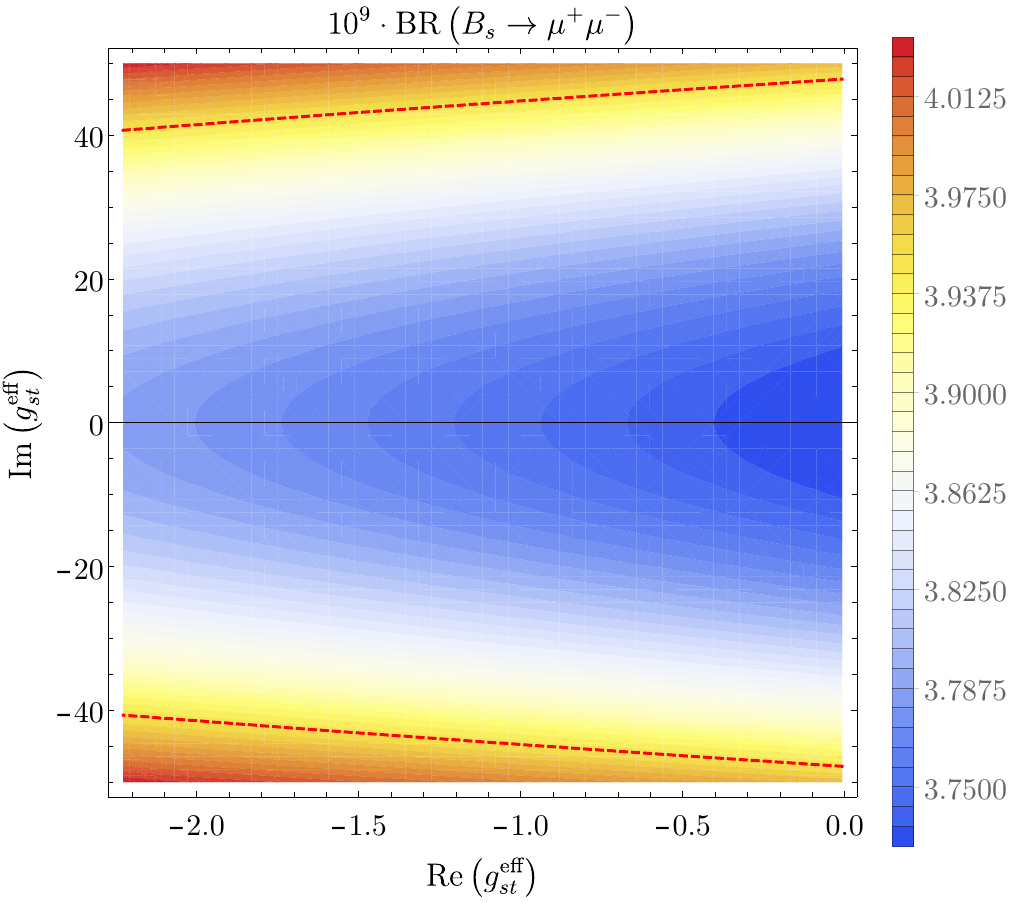}
\caption{$M_{H^\pm} = M_{H^0} = M_{A^0} = \SI{600}{\giga\eV}$, $\tan\beta = 7$.}
\end{subfigure}
\begin{subfigure}[t]{.45\textwidth}
\centering
\includegraphics[width=\textwidth]{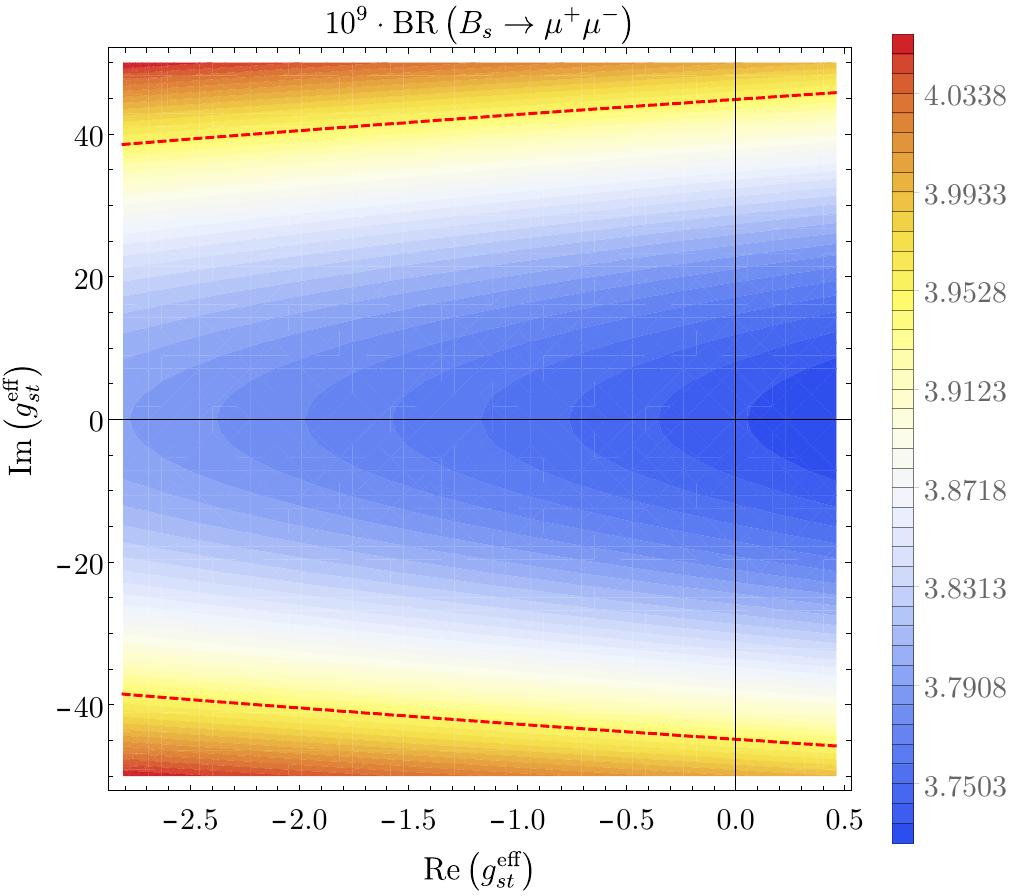}
\caption{$M_{H^\pm} = M_{A^0} = \SI{800}{\giga\eV}$, $M_{H^0} = \SI{600}{\giga\eV}$, $\tan\beta = 7$.}
\end{subfigure}
\begin{subfigure}[t]{.45\textwidth}
\centering
\includegraphics[width=\textwidth]{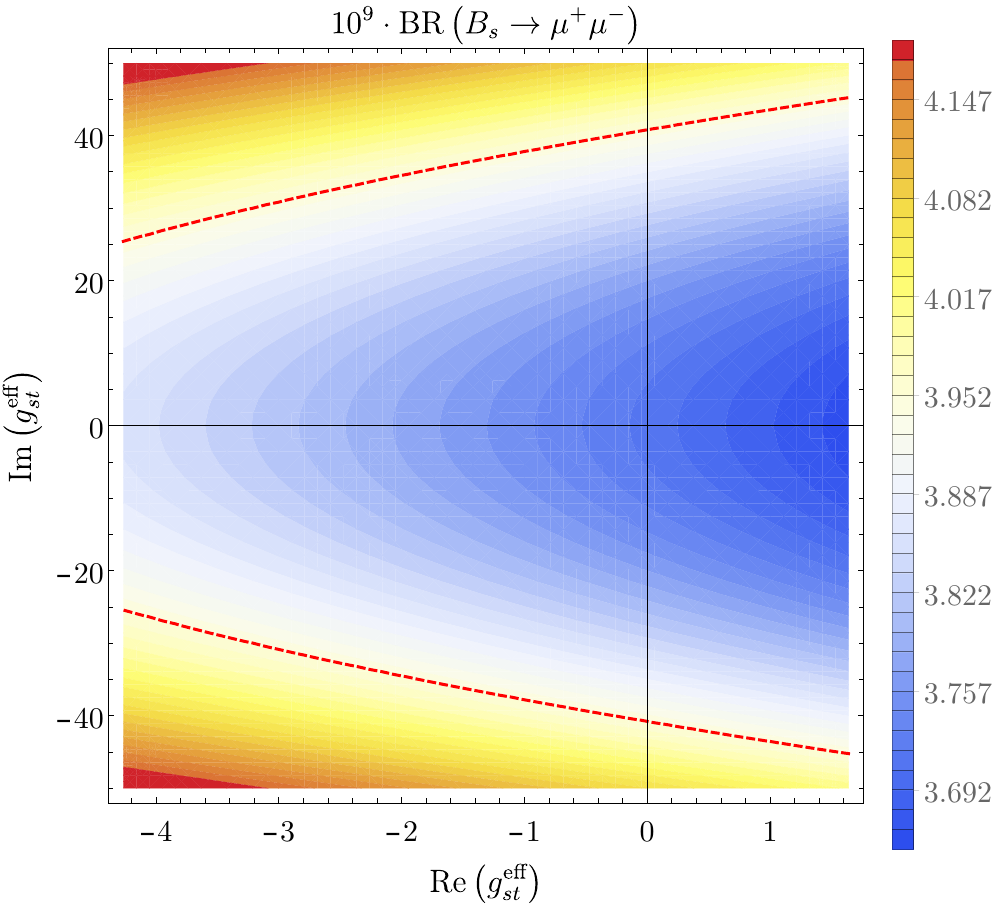}
\caption{$M_{H^\pm} = M_{H^0} = M_{A^0} = \SI{1200}{\giga\eV}$, $\tan\beta = 12$.}
\end{subfigure}
\begin{subfigure}[t]{.45\textwidth}
\centering
\includegraphics[width=\textwidth]{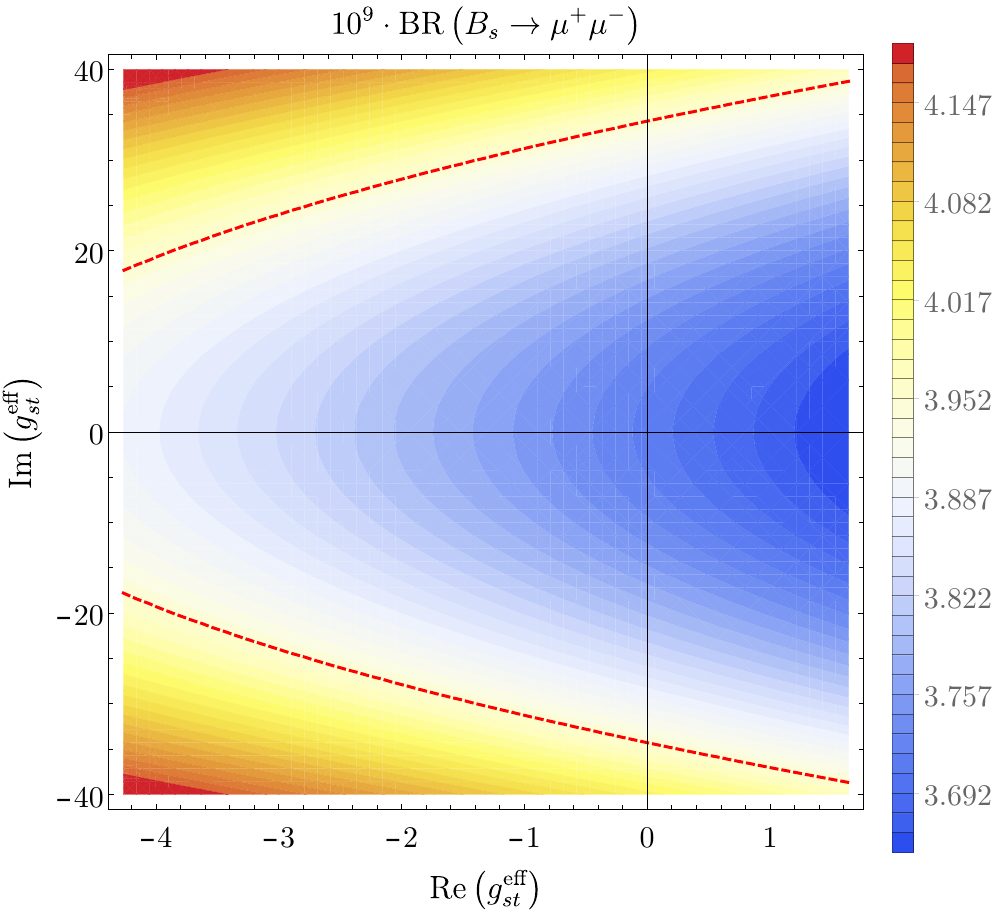}
\caption{$M_{H^\pm} = \SI{1200}{\giga\eV}$, $M_{H^0} = M_{A^0} = \SI{1100}{\giga\eV}$, $\tan\beta = 12$.}
\end{subfigure}
\begin{subfigure}[t]{.45\textwidth}
\centering
\includegraphics[width=\textwidth]{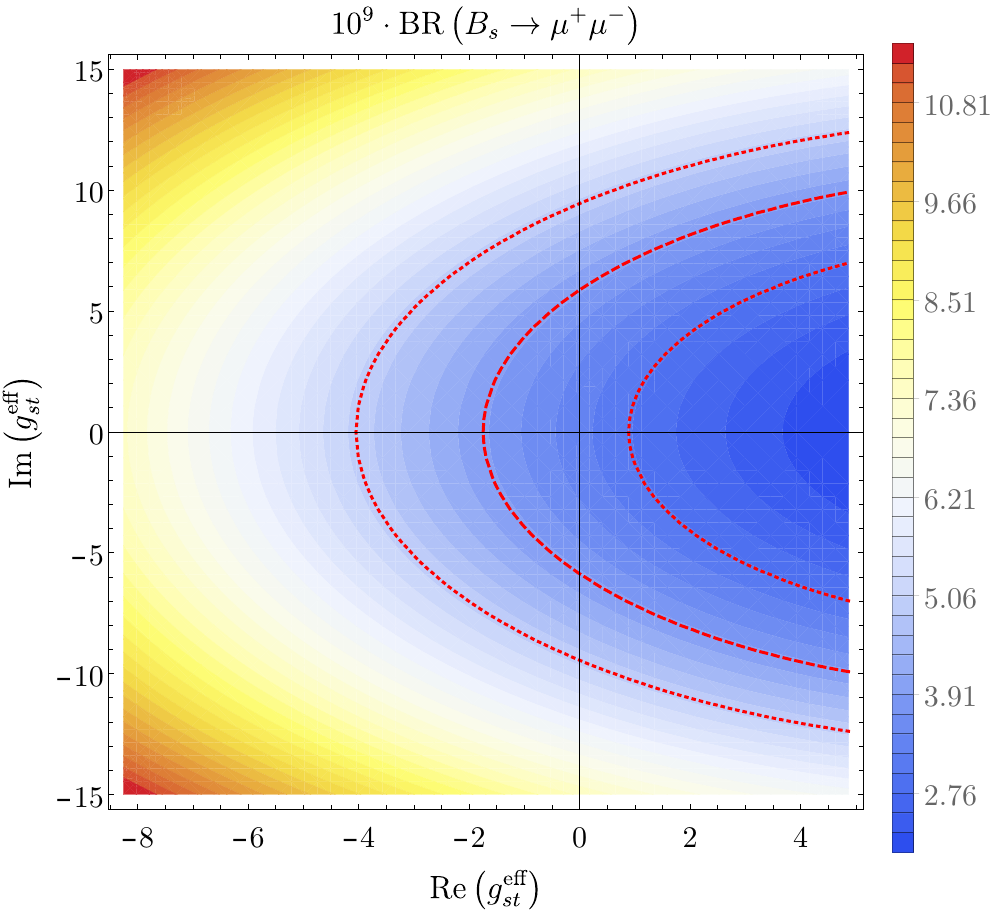}
\caption{$M_{H^\pm} = M_{H^0} = M_{A^0} = \SI{2000}{\giga\eV}$, $\tan\beta = 60$.}
\end{subfigure}
\begin{subfigure}[t]{.45\textwidth}
\centering
\includegraphics[width=\textwidth]{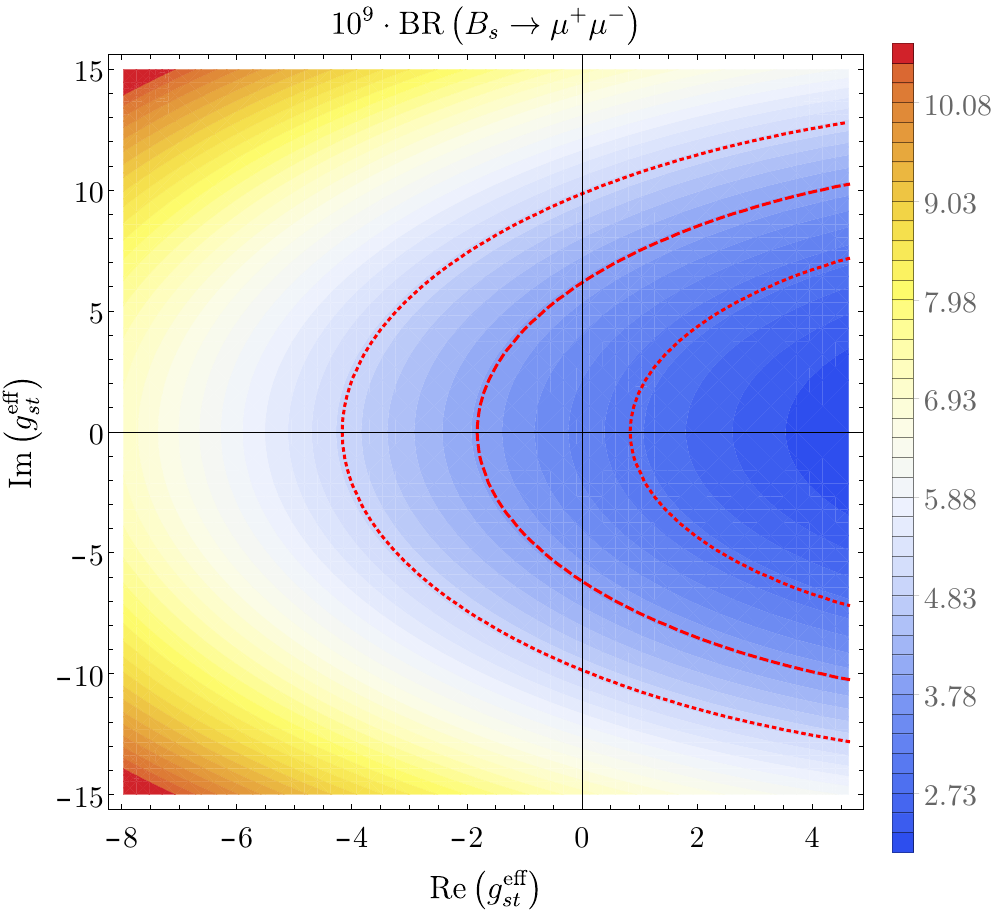}
\caption{$M_{H^\pm} = \SI{1950}{\giga\eV}$, $M_{H^0} = \SI{2050}{\giga\eV}$, $M_{A^0} = \SI{2000}{\giga\eV}$, $\tan\beta = 60$.}
\end{subfigure}
\caption{Same as \fig{fig:phenoplots1} for 
  the CMS branching ratio measurement~\cite{CMS:2022dbz}.}\label{fig:phenoplots2}
~\\[-3mm]\hrule
\end{figure}

In \fig{fig:phenoplots1} and \fig{fig:phenoplots2} we illustrate the
dependence of the branching ratio
  $\mathcal{B}\left(B_s \to \mu^+ \mu^- \right)$ on the
flavour-changing Yukawa couplings for some exemplary numerical values.
We choose to show our results separately for the
    LHCb~\cite{LHCb:2021vsc,LHCb:2021awg} and CMS~\cite{CMS:2022dbz}
    measurements, which lie on different sides of the SM
    prediction calculated with the value $|V_{cb}|$ taken from inclusive decays.
    The
    interval allowed on the horizontal axis (the real part
    $\mathrm{Re}\left(g_{st}^{\mathrm{eff}}\right)$) is for each choice
    of Higgs masses determined by the range allowed by 
    $b \to s \gamma$.  Recall that $g_{st}^{\mathrm{eff}}$
    carries a factor of $\tan\beta$, which needs to be taken into
    account if the bounds on $g_{st}^{\mathrm{eff}}$ were to be
    converted into direct bounds on $g_{ct}$. A priori
      generic $\leq {\cal O} (0.1)$ values of $g_{ct}$ could make
      $g_{st}^{\mathrm{eff}}$ in \eq{gsteff} as large as ${\cal O}(50)$. 
      For the considered Higgs masses such
        large values of $|g_{st}^{\mathrm{eff}}|$ are forbidden by
       $b\to s\gamma$ and the range for  $ \mathrm{Im}
       \left(g_{\mathrm{eff},bs\gamma} \right)$ shown on the $y$ axis in
     plots (a) to (d) only serves the purpose to show the constraint
     from  $\mathcal{B}\left(B_s \to \mu^+ \mu^- \right)$. 
     Low values for Higgs masses enforce small values for $\tan\beta$
     from the LHC searches, but for these scenarios $b \to s \gamma$
     forbids any measurable effect in $\mathcal{B}\left(B_s \to \mu^+
       \mu^- \right)$. This situation changes if one considers large
     values for $\tan\beta$ and the Higgs masses, because
     $\mathcal{B}\left(B_s \to \mu^+
       \mu^- \right)$ grows faster with $\tan\beta$ than 
      $\mathcal{B}(B\to X_s \gamma) $, see plots (e) and (f).
We find that in the considered scenarios the bounds from
$\mathcal{B}(B\to X_s \gamma)$ are stronger than those from
$\Delta M_{B_s}$, which we always find in the band allowed by
the current theoretical uncertainty. This situation changes if we go to
even larger Higgs masses, permitting larger values of
$|g_{st}^{\mathrm{eff}}|$ in $\mathcal{B}(B\to X_s \gamma)$. The
quadratic dependence of $\Delta M_{B_s}$ on $g_{st}^{\mathrm{eff}}$
makes $\Delta M_{B_s}$ a good probe of the parameter region in which
both the Higgs masses and $g_{st}^{\mathrm{eff}}$ are large.

In order to assess the quality of the perturbative expansions, we
  discuss the size of the new couplings here. The above-mentioned value
$g_{st}^{\mathrm{eff}}\sim 50$ for the product of couplings entering
$b\to s\gamma$ is large, because in \eq{gsteff}
$g_{st}^{\mathrm{eff}}$ is normalised to the type-II result which is
suppressed by $\cot \beta$. I.e.\ in the three-spurion model this
suppression is offset, which gives rise to a different phenomenology.
Including the prefactor of \eq{gsteff} results in the product of
couplings of order $0.8 |V_{tb}V_{ts}|$ for $g_{st}^{\mathrm{eff}} = 50$, so that even for
$g_{st}^{\mathrm{eff}}\gtrsim 50$ perturbation theory is well-behaved
and two-loop corrections involving more powers of $g_{jt}$, $j=u,c,t$ are tiny.

We do not study the case of sizable imaginary parts of the new FCNC
  couplings here. These imaginary parts impact 
    CP asymmetries such as 
    $A_{CP}^{\rm mix}(B_s\to J/\psi \phi)$, which will be investigated in a follow-up
    paper.   All numerical SM input parameters are given in
  \tab{tab:numinput}. Note that we do not take into account
  uncertainties in the input parameters as the dependence of the
  branching ratio is much weaker than the dependence due to the
  variation of flavour-changing Yukawa couplings.

  We stress here that the presented calculation equally applies to
  $B_d \to \mu^+ \mu^-$ with the change $V_{ts}\to V_{td}$,
    but this is not true anymore when
  considering the constraint from $B_d - \bar{B}_d$ mixing because
  of the additional dependence on the light quark mass. Since
  $m_d$ is negligibly small, the $B_d - \bar{B}_d$ mixing
    amplitude becomes insensitive to the effective $\bar{b} d A^0$ and
  $\bar{b} d H^0$ couplings in the limit of large and degenerate Higgs-boson masses. 

\begin{table}
\centering
\begin{tabular}{c|cc}
\toprule
\text{parameter} & \text{numerical value} & \text{reference} \\
\midrule
$M_{B_s}$ & \SI{5.367}{\giga\eV} & \cite{LHCb:2019rmd} \\
$m_b$ & \SI{4.18}{\giga\eV} & \cite{Zyla:2020zbs} \\
$m_s$ & \SI{0.093}{\giga\eV} & \cite{Zyla:2020zbs} \\
$m_\mu$ & \SI{0.106}{\giga\eV} & \cite{Mohr:2015ccw} \\
$\overline{m}_t \left(\overline{m}_t\right)$ & \SI{162.622}{\giga\eV} & \cite{Zyla:2020zbs} \\
$\alpha_s \left(M_Z = \SI{91.1876}{\giga\eV} \right)$ & $0.1179$ & \cite{Zyla:2020zbs} \\
$G_F$ & \SI{1.166e-5}{\giga\eV^{-2}} & \cite{MuLan:2010shf} \\
$M_W$ & \SI{80.37}{\giga\eV} & \cite{ATLAS:2017rzl} \\
$f_{B_s}$ & \SI{0.230}{\giga\eV} & \cite{Aoki:2021kgd} \\
$\Gamma_{H}^s$ & $\left(\SI{1.616}{\pico\second}\right)^{-1}$ & HFLAV \\
$\Gamma_{L}^s$ & $\left(\SI{1.427}{\pico\second}\right)^{-1}$ & HFLAV \\
\midrule
$\abs{V_{ts}}$ & $0.041$ & CKMfitter \\
$\abs{V_{tb}}$ & $0.999$ & CKMfitter \\
$\abs{V_{cs}}$ & $0.974$ & CKMfitter \\
$\abs{V_{cb}}$ & $0.041$ & CKMfitter \\
\midrule
$\mathcal{B} \left(B_s \to \mu^+ \mu^- \right)_{\mathrm{SM}}$ & \SI{3.65(23)e-9}{} & \cite{Bobeth:2013uxa} \\
$\mathcal{B} \left(B_s \to \mu^+ \mu^- \right)_{\mathrm{LHCb}}$ & $3.09_{-0.44}^{+0.48} \times 10^{-9}$ & \cite{LHCb:2021vsc,LHCb:2021awg} \\
$\mathcal{B} \left(B_s \to \mu^+ \mu^- \right)_{\mathrm{CMS}}$ & $3.95_{-0.47}^{+0.52} \times 10^{-9}$ & \cite{CMS:2022dbz} \\
\bottomrule
\end{tabular}
\caption{Numerical input used for the phenomenological analysis. For the QCD running of the top-quark mass and the renormalisation scale, we have used \texttt{RunDec}~\cite{Chetyrkin:2000yt,Herren:2017osy}. The numerical values of the CKM matrix elements have been taken from the updates provided on the \href{http://ckmfitter.in2p3.fr/}{CKMfitter} ~\cite{Charles:2004jd} web page. The numerical values of the $B_s$ decay widths have been taken from the online updates provided at the \href{https://hflav.web.cern.ch/}{HFLAV} web page~\cite{HFLAV:2019otj}.}\label{tab:numinput}
~\\[-3mm]\hrule
\end{table}

%- }}}
%- {{{ Summary:

\section{Summary}\label{sec:summary}
We have presented a 2HDM with three flavour-breaking spurions 
in the quark Yukawa sector. The model contains the established type-I
and type-II models as limiting cases and otherwise permits large FCNC
couplings in the up-type sector while naturally suppressing FCNC effects
in down sector, as required by data. Despite the large number of
parameters the model makes characteristic predictions, such as
correlations between $b\to s \mu^+\mu^-$, $b\to s \gamma$, and
$B_s - \bar B_s$ mixing, all of which involve the same combination
$g_{st}^{\mathrm{eff}}$  of fundamental parameters. Also all couplings
to right-handed down-type quarks are proportional to the quark masses as
in the type-II model.

We have studied in detail the rare decay $B_s \to \mu^+ \mu^-$,
  calculated the Wilson coefficients of the effective operators $Q_S$
  and $Q_P$, and demonstrated the consistency of the model by showing
  that the UV counterterms follow the pattern of the spurion expansion.
  Next we have calculated next-to-leading order (two-loop) QCD
  corrections to this process to (i) verify that higher-order QCD
  corrections can be correctly included (e.g.\ all UV divergences could
  be renormalised in the usual way plus countertems for our new
  couplings) and (ii) tame the sizable renormalisation-scale dependence
  of the Yukawa couplings.  Then we have studied the phenomenology of an
  FCNC coupling $g_{ct}$ of the heavy neutral Higgs bosons to top and
  charm quarks. We have found that---contrary to expectation---the 
  dominant contribution to the loop-induced $\bar s_L b_R A^{0}$ and
  $\bar s_L b_R H^0$ couplings do not come from vertex diagrams with the
  neutral Higgs coupling to the internal top-charm line, but from
  charged-Higgs couplings which inherit the dependence on $g_{ct}$
  through SU(2) symmetry. The corresponding diagram (FCNC self-energy
  with the neutral Higgs attached to the $b$ line) is enhanced by a
  factor of $\tan\beta$ compared to the vertex diagram, resulting in a
   $\mathcal{O}\left(\tan^3\beta\right)$ contribution to the  $B_s \to
   \mu^+ \mu^-$ amplitude which is absent in the type-II model.
   This feature makes  $B_s \to \mu^+ \mu^-$ a sensitive probe
   of the model even for Higgs masses well above the lower bounds
   found by the LHC experiments. For small Higgs masses, however,
   $b\to s\gamma$ precludes large effects in $B_s \to \mu^+ \mu^-$.
   In our model the dominant contribution
   to $B_s - \bar B_s$ mixing is naturally small due to a suppression
   factor of $m_s m_b/v^2$; nevertheless  $B_s - \bar B_s$ mixing sets
   constraints on the parameter space for very large Higgs masses.

\acknowledgments M.S.L.\ would like to thank Florian Herren, Syuhei
  Iguro, Lucas Kunz and Mustafa Tabet for helpful discussions. 
  The authors are grateful to Matthias Steinhauser for collaboration in
  an early stage of the project. This research was supported by the BMBF
grants 05H18VKCC1 and 05H21VKKBA as well as
Deutsche Forschungsgemeinschaft (DFG, German Research
Foundation)  within the Collaborative Research Center
\emph{Particle Physics Phenomenology after the Higgs Discovery (P3H)}\ 
(project no.~396021762 – TRR 257). 
All Feynman diagrams within this paper have been
produced using
\texttt{TikZ-FeynHand}~\cite{Dohse:2018vqo,Ellis:2016jkw}.

%- }}}

\bibliographystyle{JHEP}
\bibliography{literature}

\end{document}